\documentclass[10pt,journal]{IEEEtran} 

\usepackage{authblk}
\usepackage{times,color,soul,xspace,enumitem}
\usepackage{caption}
\usepackage{subcaption}
\usepackage{comment}
\usepackage{xspace}
\usepackage[inline,draft,nomargin,index]{fixme}
\usepackage{url}
\usepackage{amsmath,amssymb,amsfonts}
\usepackage{algorithmic}
\usepackage{graphicx}
\usepackage{textcomp}
\usepackage{booktabs}
\usepackage{bbm}
\usepackage{multirow}
\usepackage{xcolor}
\usepackage{color}
\usepackage{tcolorbox}
\usepackage{textcomp}
\tcbuselibrary{most}
\usepackage{cleveref}
\def\BibTeX{{\rm B\kern-.05em{\sc i\kern-.025em b}\kern-.08em
    T\kern-.1667em\lower.7ex\hbox{E}\kern-.125emX}}
\usepackage{tikz}
\newcommand*{\circled}[1]{\lower.7ex\hbox{\tikz\draw (0pt, 0pt)%
    circle (.5em) node {\makebox[.6em][c]{\small #1}};}}
\newtheorem{definition}{Definition}

\crefformat{section}{\S#2#1#3} 
\crefformat{subsection}{\S#2#1#3}
\crefformat{subsubsection}{\S#2#1#3}

\FXRegisterAuthor{giu}{agiu}{\textcolor{red}{Giu}}

\FXRegisterAuthor{atr}{aatr}{\textcolor{blue}{Atr}}

\FXRegisterAuthor{lin}{alin}{\textcolor{green}{Lin}}

\begin{document}

\title{Proof of Unlearning: Definitions and Instantiation}

\author[1,2]{Jiasi Weng+\thanks{+Work was done when the author was a Postdoc with City University of Hong Kong.}}
\author[2]{Shenglong Yao}
\author[2]{Yuefeng Du}
\author[1]{Junjie Huang}
\author[1]{Jian Weng}
\author[2]{Cong Wang}

\affil[1]{College of Cyber Security,~Jinan University}
\affil[2]{Department of Computer Science,~City University of Hong Kong}

\maketitle

\begin{abstract}
The ``Right to be Forgotten" rule in machine learning (ML) practice enables some individual data to be deleted from a trained model, as pursued by recently developed machine unlearning techniques.
  To truly comply with the rule, a natural and necessary step is to verify if the individual data are indeed deleted after unlearning. 
  Yet, previous \emph{parameter-space} verification metrics may be easily evaded by a distrustful model trainer.
  Thus, Thudi \emph{et al.} recently present a call to action on \emph{algorithm-level} verification in USENIX Security'22.

  We respond to the call, by reconsidering the unlearning problem in the scenario of machine learning as a service (MLaaS), and proposing a new definition framework for \emph{Proof of Unlearning} (PoUL) on algorithm level.
  Specifically, our PoUL definitions (\emph{i}) enforce correctness properties on both the pre and post phases of unlearning, so as to prevent the state-of-the-art forging attacks; 
  (\emph{ii}) highlight proper practicality requirements of both the prover and verifier sides with minimal invasiveness to the off-the-shelf service pipeline and computational workloads.   
  Under the definition framework, we subsequently present a trusted hardware-empowered instantiation using SGX enclave, by logically incorporating an authentication layer for tracing the data lineage with a proving layer for supporting the audit of learning. 
  We customize authenticated data structures to support large out-of-enclave storage with simple operation logic, and meanwhile, enable proving complex unlearning logic with affordable memory footprints in the enclave.
  We finally validate the feasibility of the proposed instantiation with a proof-of-concept implementation and multi-dimensional performance evaluation.
\end{abstract}

\section{Introduction}
Machine learning (ML) models deployed for prediction services are usually trained on user data, which can refer to the current machine learning as a service (MLaaS) paradigm.
%
Particularly, a data owner can authorize a service provider to train an ML model over his/her data and later offer black-box prediction services with the trained model.
But at a later time, the data owner might withdraw the authorization, \emph{i.e.}, sending a request to delete his/her data from the trained model, simply due to regret emotion~\cite{wang2011regretted, sleeper2013read}, or deterred by privacy attacks on trained models~\cite{shokri2017membership,tramer2016stealing,carlini2019secret,li2021membership}.
Such right of data deletion can be legally protected by privacy regulations~\cite{villaronga2018humans}, namely ``Right to be Forgotten'' (RTBF) which is explicitly stated by the European Union's General Data Protection Regulation (GDPR)~\cite{gdpr16eu} and the United States's California Consumer Privacy
Act (CCPA)~\cite{ccpa2018privacy}.
More specifically, the U.K.'s Information Commissioner’s Office~\cite{ico2020guidance} and the Federal Trade Commission~\cite{fede21ftc} recently clarify that complying with the deletion request requires retraining the model or deleting the model altogether.

Machine unlearning is a closely relevant concept, having a target model forget partial training data, but previous unlearning approaches~\cite{cao2015towards,bourtoule2021machine,ginart2019making,gupta2021adaptive,golatkar2020eternal,golatkar2020forgetting,liu2022right,warnecke2021machine} might fail to achieve the RTBF compliance in a distrustful setting.
First of all, the approaches often assume an honest server.
However, the powerful server side is likely to not delete user data in reality, which is commonly reported~\cite{dropbox2017file, facebook2018video, NYT2020deletion}, and is naturally wavering in users' trust~\cite{fool2trust, facebook2trust}.
%
%
Furthermore, a dishonest server can also strategically evade the verification metrics suggested by prior unlearning approaches, by launching forging attacks~\cite{thudi2021necessity,shumailov2021manipulating}, and therefore, Thudi \emph{et al.}~\cite{thudi2021necessity} call for action to audit unlearning, while leaving a blank space to be filled in.
Last but not least, the nature of a black-box service manner might motivate the server to maliciously fork multiple models, which may invalidate previous unlearning approaches as well as existing black-box verified methods~\cite{liu2020have, huang2021mathsf, sommer2020towards}.
For instance, the server can fork an arbitrary model (not the target model in question), and claim having deleted data from the forking model, while still offering prediction services using the target model which never deletes data at all.\looseness=-1
%

%
In light of the dishonest server, which can arbitrarily deviate from prior unlearning approaches, this work studies how to truly implement Proof of Unlearning (PoUL) for pursuing the RTBF compliance with \emph{end-to-end} assurances, \emph{i.e.}, closed-loop enforcement starting from pre-learning and prediction to unlearning and post-prediction.

\noindent\textbf{Unlearning problem.}
We clarify the unlearning problem specific for the off-the-shelf ML pipeline in MLaaS.
Starting from a trained \emph{target model} offering prediction services, when a data owner requests to delete a data point, the server should execute an unlearning process on the target model, yielding a newly predictive model which fully eliminates the effect of the data point in accordance to the unlearning goal of previous efforts~\cite{cao2015towards, gupta2021adaptive, chen2022recommendation, chen2021graph, bourtoule2021machine}.

But the server may behave dishonestly, which may forge an incorrect target model, or run an incorrect unlearning process or fork an inconsistent model for new predictions.
We next define our PoUL to prevent such misbehavior with respect to a data point deletion.\looseness=-1

\noindent\textbf{Definitions of PoUL.} 
Here involves two sequential phases aligned with the black-box service nature:
(\emph{i}) Setup phase. the server proves that a target model in question indeed learns the data point.
%
(\emph{ii}) Deletion phase. the server proves that a newly predictive model is yielded by a correct unlearning process which indeed removes the data point from the previous target model.
%
%
Within the definition scope, we require that the server should simultaneously assure the correctness of the target model, the unlearning process and the newly predictive model, such that the data owner (as a honest verifier) can be convinced of the fulfillment of his/her unlearning problem.\looseness=-1
%

PoUL is different from a recent excellent art~\cite{jia2021proof}, \emph{proof-of-learning (PoL)}, which facilitates proofs for a learning process based on the idea of re-execution.
Specifically, PoL offers a verifier a document on learning trajectory for convincing him of the truth that the learning trajectory is used to generate a particular model with overwhelming probability.
While the PoL is easy to understand and implement, it cannot be extended to implement the PoUL, due to different problem statements and threat models, figured out by Thudi \emph{et al.}~\cite{thudi2021necessity}.
%
Recent attack examples~\cite{zhang2021adversarial} can efficiently forge learning trajectory to generate a fake but valid PoL proof, which further demonstrates the difficulty of implementing the PoUL.
%
Lastly, our PoUL especially emphasizes that subsequent prediction services should be offered by an unlearned model as expected, which is not considered by the PoL.\looseness=-1

%

To pursue practice-oriented solutions, we subsequently highlight practicality requirements needed for our PoUL:
(a) \emph{Enabling generic models.} The unlearning problem in MLaaS can happen in various models, which requires proof techniques to support generic computations involved by various model architectures.
(b) \emph{Limited invasiveness.} Proof techniques should be maximally compatible with the underlying learning algorithms, so as to maintain the original ML services quality.
(c) \emph{Minimal overhead.} Standing on the already workload-intensive ML pipeline, additional proving workloads should be as affordable as possible.
Besides the server-side requirements, PoUL should also meet the following verifier-side requirements, such that the data owner can enjoy predictive services nearly as usual beyond verifying unlearning.
(d) \emph{Concise proof.} Generated proofs should be short enough to ease the storage cost of the data owner. 
(e) \emph{Efficient verification.} Proofs should also be cheaply verified considering a usually thin verifier.

\vspace{-10pt}
\subsection{Solution Overview}
We now present our solution roadmap to implement PoUL, towards complying with the deletion obligation.
We observe that PoUL relates to proof-based verifiable computation (VC)~\cite{walfish2015verifying}, especially a line of ``authenticate first and prove later" proof systems.
Simply speaking, a data owner can authorize her data to the server, along with a data digest generated by authenticated data structures (ADS)~\cite{tamassia2003authenticated}, and then keep track of particular zero-knowledge proofs issued by the server, with respect to the correctness of a target model, an unlearning process and a newly predictive model, which are not transparent to the data owner.
Yet, aware of the mentioned practicality requirements, the line of pure crypto-assisted proof systems~\cite{walfish2015verifying} are not the preferred choices (see more in Appendix~\ref{app:tec_moti}).

From another perspective, trusted execution environments (TEEs) are quickly becoming trusted computing services offered by dominant service providers, \emph{e.g.}, Alibaba Cloud~\cite{alibaba}, IBM Cloud~\cite{ibm} and Microsoft Azure~\cite{azure}, and many application efforts~\cite{le2019sgx, duan2019lightbox, gao2021teekap} demonstrate that TEEs can assist the providers in performing with
obligation compliance, \emph{e.g.}, enforcing data usage with GDPR compliance~\cite{singh2021enclaves, russinovich2021toward}.
Therefore, we are naturally encouraged to instantiate PoUL with the TEEs-backed offerings, and thereby enable native implementation and practical deployment on the server side.
%
%
Also, we admit that recent TEEs-empowered ML works~\cite{tramer2018slalom, lee2019occlumency, hanzlik2021mlcapsule, tople2018privado, ohrimenko2016oblivious, hunt2018chiron, hynes2018efficient, yuhala2021plinius, zhang2020enabling, ng2021goten, hashemi2021darknight, asvadishirehjini2022ginn} may help mitigate a certain issue within our PoUL, but we need new and holistic designs tailored for proving end-to-end RTBF compliance in MLaaS.\looseness=-1

\noindent\textbf{Our unlearning setting.}
We begin with Bourtoule \emph{et al.}'s efficient retraining-based unlearning framework~\cite{bourtoule2021machine}, towards enabling complete deletion~\cite{greengard2022can}.
Specifically, the server trains multiple chunked submodels in isolation over disjoint shards of data, and meanwhile, each chunked submodel is incrementally learned with non-overlapping slices of data points (named sliced data hereafter); when one data point is deleted, \emph{only particular submodels exactly trained on or impacted by the data point need to be retrained}, which is therefore faster than retraining a complete model from scratch.\looseness=-1

\noindent\textbf{TEEs-capable PoUL implementation.}
Standing on the unlearning setting, we incorporate ADS with Intel Software Guard Extensions (SGX)~\cite{Intel15file, costan2016intel} on the server side, to issue proofs in the Setup and Deletion phases previously described.
Our incorporation implies (\emph{i}) \emph{an authentication layer} for tracing which sliced data are learned or unlearned, and which chunked submodels are retrained or used to offer predictions; and (\emph{ii}) \emph{a proving layer} for attesting the execution correctness of the learning or prediction processes that exactly use particular authenticated data or submodels.\looseness=-1

However, it is challenging to implement the two-layer incorporation, due to the incompatible features between our unlearning setting and Intel SGX.

\noindent\emph{\textbf{Challenge (i): Memory-Efficient Authentication.}} In the authentication layer, we require tracking which sliced data impact which chunked submodels, and supporting authenticated updates on the impacted submodels when a data point is deleted.
But both data and submodels are often large-size and the unlearning problem might cause a large number of submodels to be updated, which inevitably incurs large memory footprints far beyond the original memory limitation of SGX\footnote{It is about $94$~MB available to applications in a widely adopted version, and a recent release~\cite{intel2020sheet} supports $188$~MB.}.
Hence, we provide a memory-efficient authentication design tailored for the traceability from the sliced data to the updatable submodels while protecting the integrity.\looseness=-1

\noindent\emph{\textbf{Challenge (ii): Proof of Stateful Computation and Fast Verification.}} In the proving layer, the computations involving a chain of incremental retraining processes to update affected submodels, and subsequent prediction processes using the newly retrained model should be attested, so that a verifier is convinced of the fulfillment of data deletion.
Yet, the computations are stateful, since the incremental retraining and prediction processes require previously generated states, \emph{e.g.}, submodels, which is in conflict with the SGX design of primarily protecting stateless computations.
We extend the trust outside the SGX's protected memory to securely save previous states, by typically letting the protected memory retain unique randomnesses (named \textsf{seed}) respective to each submodel for subsequent integrity checking.
To reduce verification cost, we adopt a self-verification strategy.
Concretely, we enable the verifier to only assert that an appointed submodel (\emph{i.e.}, a final one) matching a newly produced digest yields new predictions on a given test data, while the authenticity of all retrained submodels prior to the appointed submodel is verified by the enclave itself.
Furthermore, in light of the need for monitoring the correctness of subsequent prediction services, an additional SGX-enabled auditor is considered.

In summary, we make three-fold contributions as following:
\begin{itemize}
    \item We propose a new two-phase definition framework for achieving PoUL, which adapts to Bourtoule \emph{et al.}'s generic unlearning algorithm (in IEEE S\&P'21), and meanwhile, prevents recent forging attacks (in USENIX Security'22). 
    \item Under the definition framework, we present an SGX-capable solution with newly customized designs, by logically integrating an authenticated layer for tracing the lineage of training data with a proving layer for auditing the correctness of model training. 
    \item We give a proof-of-concept implementation for the SGX-capable PoUL, and evaluate the multifaceted performance in terms of storage cost, deletion time, learning/unlearning time and trained accuracy.
    The implementation code is available in https://github.com/James-yaoshenglong/unlearning-TEE.
\end{itemize}

\section{Background}\label{sec:background}

\subsection{Machine Unlearning}\label{subsec:unlearning}
Machine unlearning starts from a concept of removing a/some training data from an already trained model~\cite{cao2015towards}.
According to the different understanding of the concept, existing approaches can be roughly  classified into two groups, including \emph{approximate unlearning} achieving data deletion from parameter level~\cite{golatkar2020eternal,golatkar2020forgetting, ginart2019making, sekhari2021remember} and \emph{exact unlearning} achieving data deletion from algorithm level~\cite{cao2015towards,bourtoule2021machine,gupta2021adaptive,chen2022recommendation, chen2021graph}.
%
%
The essential difference between the two groups is clarified by the Definition \ref{def:MLunlearning}.\looseness=-1

\begin{definition}\label{def:MLunlearning}
Let $D=\{d_i\}\cup d_u, i\in \{1,\dots,N\}\backslash u$ be a collection of $N$ training samples.
Let $D_{-u}$ denote a collection of $N-1$ training samples, namely $D\backslash d_u$.
Let $\mathcal{D}_M$ be the distribution of a model that has ever been trained over $D$ and then unlearned $d_u$ via a machine unlearning algorithm \textsf{R}.
Let $\mathcal{D}_M^{'}$ be the distribution of a model trained over $D_{-u}$.
Then denote \textsf{R} as an \textbf{approximate unlearning} algorithm, if the distribution $\mathcal{D}_M$ is approximately equal to $\mathcal{D}_M^{'}$.
Otherwise, denote \textsf{R} as an \textbf{exact unlearning} algorithm, if the distribution $\mathcal{D}_M$ is exactly equal to $\mathcal{D}_M^{'}$.
\end{definition}

Recently, \emph{the \text{SISA} framework}~\cite{bourtoule2021machine} provides an exact unlearning method towards generic ML models, which fully erases the effect of deleted data.
It is faster than retraining from scratch by trading storage for efficiency, and is followed by many promising unlearning work~\cite{gupta2021adaptive, chen2022recommendation, chen2021graph}. 

\begin{figure}[ht]
\small
    \centering
    \includegraphics[width=0.95\linewidth]{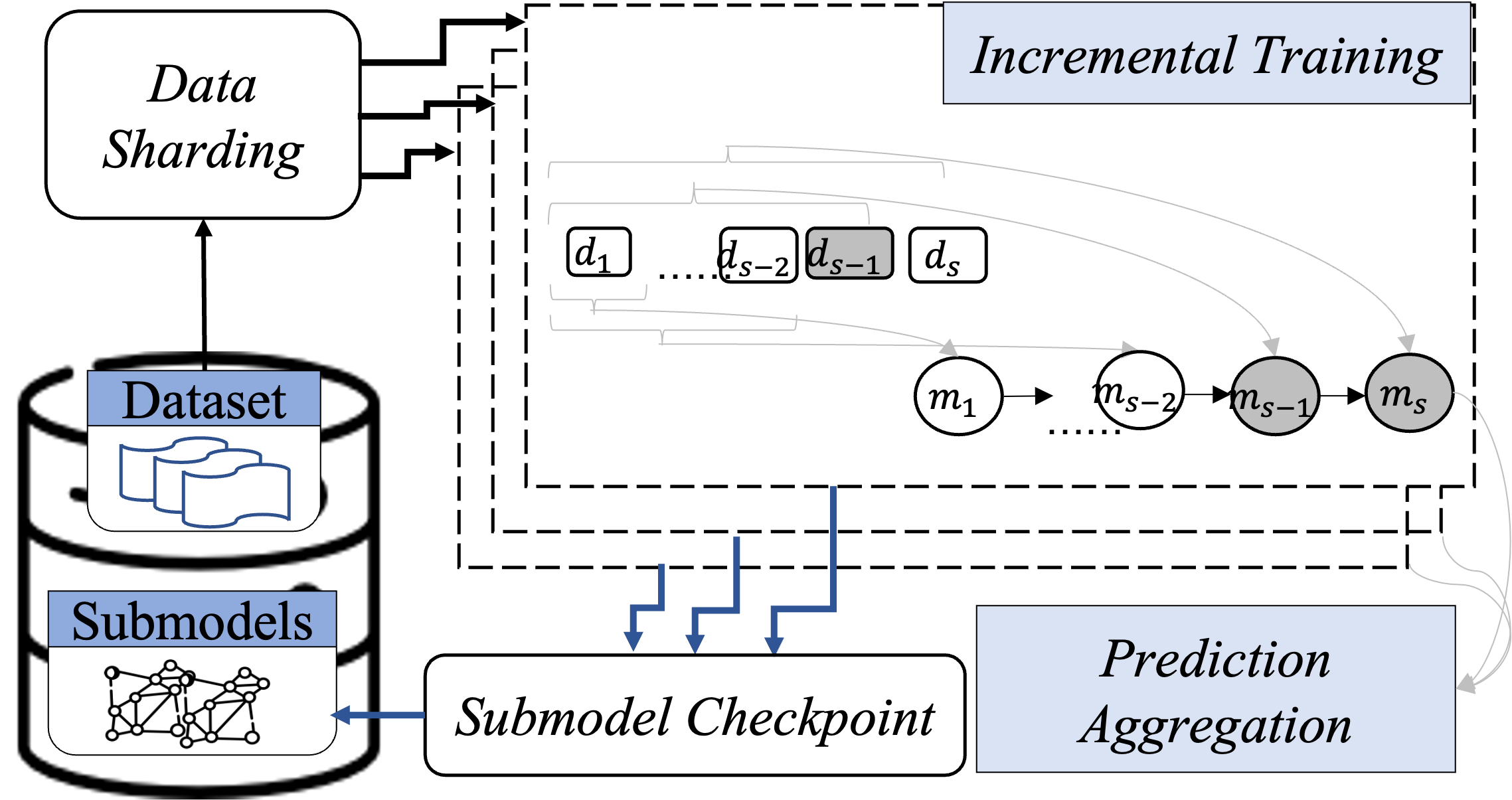}
    \caption{The server-side ML pipeline under the SISA framework. Multiple dashed rectangles represent that different \emph{Shards} of data are used for incrementally training constituent models in an \emph{Isolated} manner. Within one shard, the training data is further \emph{Sliced} into disjoint data slices, where the data slice in grey (\emph{i.e.}, $d_{s-1}$) influences or impacts the intermediate submodels in grey (\emph{i.e.}, $m_{s-1}$ and $m_{s}$). All constituent models (the last submodels) of individual shards produce predictions that are \emph{Aggregated} into final predictions.}
    \label{fig:pipeline}
    \vspace{-12pt}
\end{figure}

We next introduce a server-side ML pipeline under the SISA framework, as shown in Fig.~\ref{fig:pipeline}.
Suppose the server-side pipeline runs over the dataset authorized by a data owner and generates learned models, along with model checkpoints.
At a later time, it complies with a deletion request of the data owner via unlearning.
We introduce the pipeline with two parts: A) pre-learning and prediction stages, as well as B) unlearning and post-prediction stages.
We denote $N=\{1,\dots,n\}$ and $S=\{1,\dots,s\}$.

\noindent\textbf{A. Pre-learning and Prediction Stages} There are four major operations. 
(a) \emph{Shard.} In the first place, the data owner's dataset is divided into $n$ non-overlapping data shards, denoted as $D_{j\in[N]}$;
(b) \emph{Isolation.} Subsequently, stochastic gradient descent (SGD)-based training algorithm is applied to train a constituent model on each shard of data in isolation;
(c) \emph{Slice.} Inside each shard, the data $D_j$ is further sliced into $s$ disjoint data slices, represented by $D_j=\{d_{j,i}\}_{i\in[S],j\in[N]}$, such that data slices are incrementally added for learning, and the produced submodels $\{m_{i}\}_{i\in[S]}$ are saved during learning;
(d) \emph{Aggregation.} The final prediction is obtained by aggregating the predictions provided by the constituent models $m_{s,j}, j\in[N]$ of all shards. 
Herein, a data slice can contain a small set of \emph{non-overlapping} data points.
Note that the impact of each data point is restricted on relatively small-size submodels, rather than an entire model.
%
%

\noindent\textbf{B. Unlearning and Post-prediction Stages} On receiving the data owner's request of deleting one data point $d_u\in D$, the following steps are executed: 
(a) find which shards this data point $d_u$ belongs to and which submodels it influences; 
(b) locate the shard and the slice associated to $d_u$, and delete the $d_u$ from the slice and the submodels influenced by $d_u$ within the shard; 
(c) re-execute the incremental training processes with the remaining data slices as pre-training.

Now we show a concrete example.
We suppose the $d_u$ that will be deleted falls in the last but one slice $d_{s-1}$ of Fig.~\ref{fig:pipeline} and the submodels $m_{s-1}$ and $m_s$ are influenced by the $d_u$.
Guided by the above steps, the $m_{s-1}$ and $m_s$ will be deleted, and subsequently, new $m_{s-1}$ and $m_s$ are relearned over the remaining data $\{d_1, \cdots, d_{s-2}, d_{s-1}\backslash d_u\}$ and $\{d_1, \cdots,$ $d_{s-2}, d_{s-1}\backslash d_u, d_s\}$, respectively, in the same incremental manner.
Finally, new predictions are produced by newly updated models and aggregated without any effect of the deleted data point.
Notice, the server only needs to re-compute the relatively small-scale submodels affected by the $d_u$, without the need to retrieve or relearn the entire model.
As a result, this framework can be faster than retraining from scratch, varying from $2.45\times$ to $4.63\times$~\cite{bourtoule2021machine}.

\subsection{Trusted Execution Environments}
Trusted Execution Environments (TEEs) provide a hardware-protected memory environment for sealed data and shielded execution of applications in an untrusted platform.
Many off-the-shelf TEEs technologies, \emph{e.g.}, Intel SGX~\cite{Intel15file, costan2016intel}, ARM TrustZone~\cite{alves2004trustzone}, become promising to implement secure applications with minimal performance compromise, in which Intel SGX is widely adopted, and thus our work adopts SGX.

\noindent\textbf{Intel SGX Enclave} Intel SGX allows creating a secure memory region, named as \emph{enclave}, to execute application codes with confidentiality and integrity guarantees, \emph{isolated} from the outside platform which can be untrusted.
%
It also provides a \emph{remote attestation} mechanism by issuing a signature-based proof (Enhanced Privacy ID signature scheme~\cite{brickell2011enhanced}) on a requested \emph{quote}, containing the measurement of the enclave's code and data.
With the proof, a remote client can assert the authenticity of the enclave identity and the truthfulness of code execution, after it subscribes to the Intel's Attestation Service.

\noindent\textbf{Formal Functionality Modeling}
Shi \emph{et al.}~\cite{pass2017formal} formalize the functionality of an SGX enclave, involving enclave initialization, enclave operations and attestation. 
With the functionality, Tram{\`e}r \emph{et al.}~\cite{tramer2017sealed} introduce an enclave-empowered ``commit-and-prove" functionality executed by a prover equipped with a \emph{transparent} enclave and a verifier.
The transparent enclave is considered with integrity property, but with minimal confidentiality assumptions, \emph{e.g.}, random number generators and the signing key, not including the programs running in the enclave.
More detailed description refers to Appendix \ref{app:fun_sgx}.

\noindent\textbf{Attack Threats on SGX Enclave} Integrity violation attacks on the SGX enclave, such as forking, replacing, relocation and rollback attacks will be mitigated by our designs (see Section~\ref{sec:structure}).
However, we do not address existing side-channel attacks breaking integrity by stealing secret keys~\cite{chen2019sgxpectre, van2020sgaxe}, aware of recent countermeasures protecting secret-dependent memory accesses against such side-channel attacks.
Control flow attacks~\cite{abera2016c, dessouky2017fat} are also outside of our scope, since the attacks escape from the off-the-shelf static remote attestation mechanisms.
In addition, Denial-of-Service attacks~\cite{jang2017sgx}, \emph{e.g.}, shutting down the enclave applications, are not our concern.

\subsection{Data Structure for Fast Membership Testing}
We elaborate here a space-optimized and high-speed data structure, namely, cuckoo filter~\cite{fan2014cuckoo}, designed for membership query from a usually large data set with low and controllable false positive rates.
Different from Bloom filters, it supports $O(1)$ element deletion, not merely addition. 
Compared with cuckoo hash tables, it only stores short and constant-size fingerprints of elements.

Concretely, a cuckoo filter consists of an array of buckets, in which each bucket contains multiple entries, \emph{e.g.}, $4$ or $8$. A fingerprint of an element can be filled in two possible buckets. This is determined by two hash functions derived from standard cuckoo hashing~\cite{paghR04cuckoo} as introduced in the following.
For example, given a new element $e$ to be inserted and its fingerprint $f_e$ (\emph{e.g.}, truncated PRESENT ciphertext with $16$~bits as \cite{shafi2020secsched} applied), an alternative location $h_1$ of bucket $B$ is found by calculating a $64$-bit hash value $h_1=\text{Hash}(e)$.
The hash functions can adopt the CityHash~\cite{fan2014cuckoo}.
%
If it is empty, the fingerprint of this element is inserted into $B[h_1]$.
Otherwise, another location $h_2$ is found by calculating $h_2=h_1\oplus \text{Hash}(f_e)$, where the original element $e$ is not needed to be retrieved.
Next, a location displacement method is called, if $B[h_2]$ is also not empty.
Specifically, the element at $h_2$ would be replaced with $f_e$, and the element is placed in its alternative location.
This displacement process is repeated until an empty bucket is found or a displacement threshold (\emph{e.g.}, 500) is reached.\looseness=-1

\section{Problem Statement}\label{sec:problem}
We observe that a dishonest sever can arbitrarily deviate from fulfilling a data deletion request of a data owner during unlearning.
Also, there not exists an approach to enforcing the behavior of the server during unlearning.
The observation motivates us to establish trust for the data owner.
This section will clarify the threats and deletion assumptions we are concerned about, for ease of understanding our definitions of PoUL in the next section.

\subsection{Threat Scenario}
\begin{figure}[ht]
\vspace{-5pt}
\small
    \centering
    \includegraphics[width=0.8\linewidth]{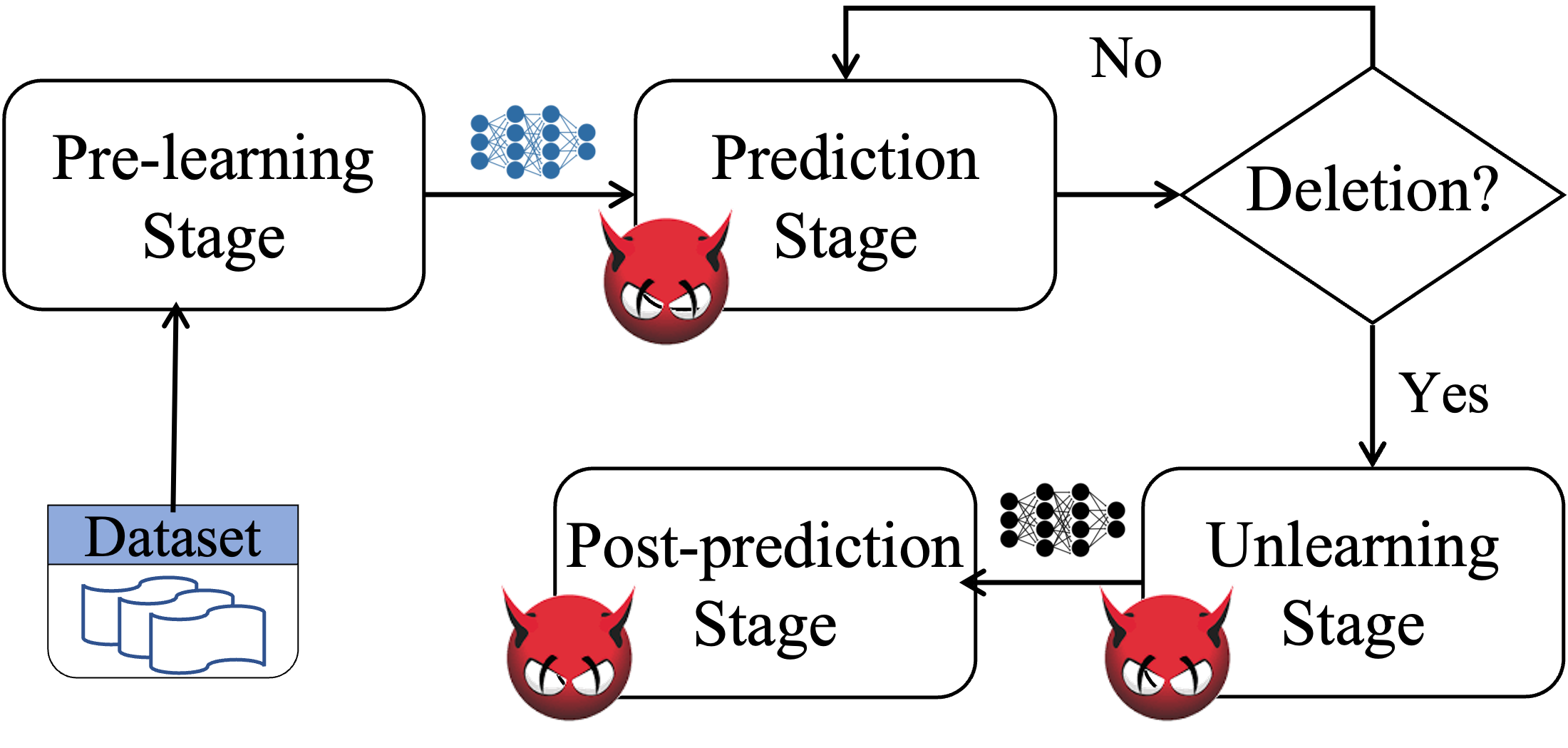}
    \caption{Three threats to the server-side ML pipeline.}
    \label{fig:overview}
    \vspace{-12pt}
\end{figure}
On top of the ML pipeline previously described, we consider the server can misbehave arbitrarily, by forking multiple models due to the non-transparent nature of the black-box service manner, or forging unlearned models with off-the-shelf forging attacks~\cite{thudi2021necessity,shumailov2021manipulating}.
We note prior auditing methods~\cite{liu2020have, huang2021mathsf, sommer2020towards, xiao2022verifi} are not concerned about the misbehaviors.
Concretely, we summarize the following three cases of misbehaviors in the pipeline assisted with Fig.~\ref{fig:overview}:
\begin{itemize}
    \item At the prediction stage, the server might substitute a \emph{correct target model} and avoid deleting data from the target model. The so-called correct target model is denoted as the model which was ever trained using the data to be deleted.
    \item At the unlearning stage, the server might not correctly execute an unlearning process.
    \item At the post-prediction stage, the server might not deploy a correct unlearned model for subsequent service.
\end{itemize}

For data owner, we consider he/she is honest throughout the process, only having a black-box access to the models.
His/Her data collected by the server for learning and test data for predictions are also considered benign.
Data confidentiality against the server is not considered.
Lastly, we let the the data owner communicate with the server via an authenticated point-to-point communication channel.\looseness=-1

\subsection{Deletion Assumptions}
We focus on the problem of deleting training data from a target model on the server site.
%
%
%
We note that data deletion from memory on physical medium addressed by previous arts~\cite{reardon2013sok, onarlioglu2018eraser} is outside of our consideration.
%
%
We are not concerned about data or model copy, simply due to monetary storage costs, and do not consider deleting copied data or model.
%
%
We assume the training data is non-overlapping similar to the SISA framework.
We focus on a single data deletion request, but our design can be extended to delete multiple data pieces (see Appendix~\ref{app:discussion}).
Relying on the SISA framework's technical characteristic, multiple data deletion requests should be assumed irrespective of the actual models (\emph{e.g.}, a never unlearned model or an unlearned model) in question~\cite{bourtoule2021machine}, and adaptive deletion~\cite{gupta2021adaptive} is out of our scope.
Additionally, recent inference attacks~\cite{chen2021machine, gao2022deletion} on unlearned models are out of consideration.\looseness=-1



\section{Definitions on Proof of Unlearning}\label{sec:definition}
%
%
%
We are ready to describe the definitions on PoUL, along with three correctness properties and five practicality goals.
Intuitively, we require the server to generate proofs of correct unlearning, and given the proofs, the data owner has confidence in asserting if the server complies with his/her deletion request or not.
%

To begin with, we denote some notations. 
We let an indexed collection of learning data $D=\{d_i\}_{i\in[N]}$, including a data point $d_u$ that will be deleted.
We denote $\textsf{M}_0$ as a public model with initial weights, and $\textsf{F}$/$\textsf{G}$ as public learning/prediction algorithms.
We let a test data $t$ and a prediction $p$.
We mark $[m_u]$ as the affected part of a model by its learning data $d_u$.
We use \textsf{SR} and \textsf{DO} to denote the server and the data owner, respectively.
We also use $\pi$ to define a proof component generated by the  \textsf{SR}.
%
%

\subsection{Definition Framework} 
There are two phases composing the definition framework for PoUL. 
In a setup phase, the \textsf{SR} offers a proof attesting that the currently predictive model is a true target model responding to a test challenge from the \textsf{DO}.
In a later deletion phase, the \textsf{SR} retains a proof of truth that the newly predictive model is a truly unlearned model from the above target model excluding the deleted data piece $d_u$, responding to a new test challenge.
Particularly, the two phases involve a chain of interactive procedures as following.

\noindent\textbf{Setup Phase}
($\textsf{S}_1$-\textsf{Initialize}) The \textsf{DO} uploads his/her data $D$ along with an authentic data digest to the \textsf{SR};
The \textsf{DO} specifies a learning algorithm $\textsf{F}$ and an initial model $\textsf{M}_0$ with the \textsf{SR};
The \textsf{SR} then learns a model $\textsf{M}_1$ from $\textsf{M}_0$ with $\textsf{F}$ taking the data $D$ as input.
Notice, the learned $\textsf{M}_1$ now is the target model we describe before, and its digest $\textsf{H}(\textsf{M}_1)$ is published.
($\textsf{S}_2$-\textsf{Challenge}) The \textsf{DO} sends a test data $t$ to the currently predictive model for prediction query.
($\textsf{S}_3$-\textsf{Prove}) The \textsf{SR} responses to the query with a prediction $p$ and a proof component $\pi$.
($\textsf{S}_4$-\textsf{Verify}) With the proof $\pi$ and prediction $p$, and the above model digest $\textsf{H}(\textsf{M}_1)$, the \textsf{DO} verifies the correctness of the prediction.
Specifically, he/she would reject it with high probability, if the prediction is not yielded by the target model $\textsf{M}_1$, denoted as a statement $p\leftarrow \textsf{G}(\textsf{M}_1, t)\wedge \textsf{M}_1\leftarrow \textsf{F}(\textsf{M}_0,D)$.

\noindent\textbf{Deletion Phase}
($\textsf{D}_1$-\textsf{Unlearning}) The \textsf{DO} sends a request of deleting the data $d_u$ from the target model $\textsf{M}_1$;
the \textsf{SR} executes an unlearning process of $\textsf{F}$ on the $\textsf{M}_1$ and $D\backslash d_u$, yielding an unlearned model $\textsf{M}_2$ whose digest $\textsf{H}(\textsf{M}_2)$ is public.
($\textsf{D}_2$-\textsf{Challenge}) At a later time, the \textsf{DO} challenges the newly predictive model with a new test data $t'$.
($\textsf{D}_3$-\textsf{Prove}) The \textsf{SR} returns a prediction $p'$, along with a new proof component $\pi'$.
($\textsf{D}_4$-\textsf{Verify}) With the proof $\pi'$ and prediction $p'$, and the digests $\textsf{H}(\textsf{M}_1)$ plus $\textsf{H}(\textsf{M}_2)$, the 
\textsf{DO} verifies the prediction correctness, and rejects it with high probability, if the prediction $p'$ is not offered by a correctly unlearned model, denoted as a statement $p' \leftarrow \textsf{G}(\textsf{M}_2, t')\wedge \textsf{M}_2\leftarrow \textsf{F}(\textsf{M}_1\backslash [m_u],D\backslash d_u)$.\looseness=-1
%
%

\noindent\textbf{Logical Components}
Essentially, there are two-layer components to be realized for supporting the setup and the deletion phases in our scoped PoUL:
\textbf{\emph{(1) Authentication layer}}:
The data owner authenticates his/her data (\emph{e.g.}, data digest) in the first place and delegates the data to the server, along with the authenticated result for tracking operations on the data in a future time, \emph{e.g.}, using or not using the data for training.
Relying on the authentication layer, the intermediate models and the final model yielded by a learning process, should also be authentically saved (and updated), such that a later unlearning process or a prediction process is ensured to use the previously authenticated models.
\textbf{\emph{(2) Proving layer}}: 
This layer works jointly with the authentication layer.
Particularly, the server convinces the data owner that a learning process, either in the $\textsf{S}_1$-$\textsf{Initialize}$ or the $\textsf{D}_1$-$\textsf{Unlearning}$ procedure, is executed as expected, with the learning data matching the previously authenticated data.
Besides learning, the server also assures the correctness of a predictive model, which exactly means that a prediction process in the $\textsf{S}_3$-$\textsf{Prove}$ procedure (resp. $\textsf{D}_3$-$\textsf{Prove}$) uses a most recent model yielded by the $\textsf{S}_1$-$\textsf{Initialize}$ procedure (resp. $\textsf{D}_1$-$\textsf{Unlearning}$).

\subsection{Correctness Properties} 
We define three correctness properties that should be satisfied by PoUL. 

\noindent$\bullet$ \emph{Target Model Correctness.}
A currently predictive model $\textsf{M}_1^{'}$ yielding a prediction $p$ respective to a test data $t$, is a correct target model $\textsf{M}_1$, if the statement $p\leftarrow \textsf{G}(\textsf{M}_1^{'}, t) \wedge \textsf{M}_1^{'}=\textsf{H}^{-1}(\textsf{M}_1) \wedge \textsf{M}_1\leftarrow \textsf{F}(\textsf{M}_0,D)$ is true, where $\textsf{G},\textsf{F}$ and $\textsf{M}_0$ are publicly known.

\noindent$\bullet$ \emph{Unlearning Correctness.} 
The unlearning process for deleting the $d_u$ from the target model $\textsf{M}_1$ is correct, if $\textsf{M}_2\leftarrow \textsf{F}(\textsf{M}_1\backslash [m_u],D\backslash d_u)$ is true, where $[m_u]$ included in $\textsf{M}_1$ is the part exactly impacted by the $d_u$.

\noindent$\bullet$ \emph{New Model Correctness.}
A newly predictive model $\textsf{M}_2^{'}$ yielding a new prediction $p'$ for a new test data $t'$, is a correct new model with respect to the deleted $d_u$, if the statement $p' \leftarrow G(\textsf{M}_2^{'}, t') \wedge \textsf{M}_2^{'}=\textsf{H}^{-1}(\textsf{M}_2)$ is true.

Specifically, in the setup phase, we require guaranteeing the target model correctness, such that the server cannot fake a target model that never learns the data owner's data, and claim unlearning from it in the later deletion phase.
In the deletion phase, unlearning correctness and new model correctness should be ensured, such that the server cannot execute an incorrect unlearning process, and cannot fork an arbitrary model for subsequent predictions while claiming the fulfillment of unlearning.
If the above correctness properties are satisfied, the data owner can be convinced that his/her deletion request is truly addressed by the server.

\subsection{Practicality Goals}\label{subsec:goal}

\noindent$\bullet$ \emph{Generic Model Supports.}
%
Data deletion requests can occur in generic ML model scenarios, as previous unlearning approaches~\cite{bourtoule2021machine} supported. 
The definition of PoUL should cover generic ML models.

\noindent$\bullet$ \emph{Limited Invasiveness.}
We desire that proof generation has little modification effect on the underlying learning and unlearning pipelines, and should not compromise model accuracy\footnote{Our paper considers single data deletion, which generally makes no obvious effect on model accuracy~\cite{villaronga2018humans}}. 

\noindent$\bullet$ \emph{Minimal Overhead.}
Proof generation should not incur additional unaffordable overhead to the already intensive workloads.

\noindent$\bullet$ \emph{Concise Proof.}
Proof size should be short, compared to giving a PoL document on training trajectory~\cite{jia2021proof}.

\noindent$\bullet$ \emph{Efficient Verification.} 
Verification cost is expected to be small and constant, faster than re-executing and validating partial training trajectory as \cite{jia2021proof}.
%

\section{Our Designs}\label{sec:design}
We present our SGX-protected designs for the PoUL on top of the SISA unlearning framework, with a partition of an authentication layer and a proving layer.
We mainly place the large authenticated storage with simple operation logic outside the enclave, and preserve complex provable execution logic but minimal storage costs in the enclave.
This section begins with two challenges of implementing PoUL, considering the server equipped with an SGX enclave.
%
%
In Section~\ref{sec:structure}, we customize data structures for authenticating learning data and intermediate submodels.
In Section~\ref{sec:proving}, we integrate the customized data structures with a proof  protocol for realizing PoUL.

\subsection{Challenges and Solutions}\label{sec:chall}
Although Intel SGX provides the integrity property that is important to our scenario problem, its inherent limitations make us encounter challenges in order to efficiently implement PoUL.
We next introduce two main challenges, with respect to the authentication layer and the proving layer we need.

\noindent(1) \emph{\textbf{Authentication layer}: How to efficiently track the lineage from authenticated learning data to learned intermediate submodels while supporting authenticated deletion?}

It is challenging due to the incompatibility of memory-constrained SGX enclave and memory-intensive workloads.
%
Combining with the SISA training process in Section~\ref{subsec:unlearning}, one possibility of implementing the authentication layer is to move data and submodels outside the enclave (\emph{e.g.}, persistent memory or disks), and meanwhile, adopt appropriate compact ADS to authenticate them as well as keep the lineage from the data to the corresponding submodels.
In particular, we may leverage two same-size Merkle hash trees (MHTs) to separately package data and the corresponding intermediate submodels, aligning with sliced indices in SISA.
Their lineage is preserved, by making the location of a data slice in one MHT consistent with that of the \emph{earliest} intermediate submodel influenced by the data slice within another MHT.
Lastly, the two entire MHTs reside out of the enclave and their constant-size roots are stored in the enclave for integrity check.
Such adoption, however, is not efficient, in light of the costs to update the two MHTs at worst-case unlearning of SISA.
That is, ($\textsf{op}_1$) deleting a data piece from the first data slice of a shard, and simultaneously, ($\textsf{op}_2$) updating a sequential of submodels affected by the data piece.
We suppose $n_d$ and $n_m$ leaves in the two MHTs.
Then, $\textsf{op}_1$ consumes $n_d \times \text{log}~n_d + \text{log}~n_d$ hash evaluations for checking the integrity of $n_d$ data slices and updating the first data slice.
$\textsf{op}_2$ needs $2\times n_m \times \text{log}~n_m$ hash evaluations for integrity check and update on the $n_m$ submodels. 
To assure correctness, both $\textsf{op}_1$ and $\textsf{op}_2$ also need to be operated in the secure enclave, and thus unfortunately it is not cost-efficient.

\textbf{\emph{Solution.}} We reconsider that the necessity of adopting the
above ADS is aimed to track the data lineage correlated with the intermediate submodels while supporting dynamic updates. 
To relieve the strict configuration on the ADS and avoid tree traversals, we let the enclave itself achieve the data lineage tracking with correctness guarantees, by designing highly memory-efficient data structures (\emph{i.e.}, compact index list and filter) and cheap pointers interlinking data slices with correlated submodels out of the enclave. 
Besides, data and submodels are authentically saved outside the enclave, and are protected against potential integrity attacks, with our strategical data structure designs in the enclave.

\noindent(2) \emph{\textbf{Proving layer}: How to enable efficient verifiability on the correctness of predictive models and incremental learning which involve stateful computations, using previously authenticated submodels and learning data?}\looseness=-1

This challenge is caused by the conflict between our essential requirements on attesting stateful computations, \emph{e.g.}, multiple incremental learning processes, and the SGX's protection primarily on stateless in-memory computations, which cannot allow a low and constant verification cost.
To be specific, we require the enclave to execute each incremental learning process, taking as input a most recent submodel and newly incremental learning data, and finally yield a new submodel.
A verifier then can assert the correctness of the new submodel by verifying a signature.
Due to the incremental nature, the verifier who asserts the correctness of a predictive model (\emph{i.e.}, the final submodel) needs to assert that all submodels learned prior to this final submodel are correct.
For example, when the data owner wants to verify if a newly predictive model is not trained on her data piece originally falling in the most beginning data slice, she needs to verify the correctness starting from the most beginning submodel till the final one.
Such verification overhead linearly depends on the number of submodels affected by the data slice.

\textbf{\emph{Solution.}} We enable a constant verification cost by letting the verifier only assert the correctness of the final submodel, but it in turn requires extending the SGX's protection to stateful computations.
Therefore, we consider a self-verification method.
That is, before verifying the final model, the authenticity of all prior submodels is verified by the enclave itself, by securely retaining unique randomnesses (named \textsf{seed}) respective to each submodel.
To further assert the correctness of subsequent predictions, we additionally introduce a trusted enclave, serving as an \emph{auditing enclave}.
The auditing enclave is responsible to send challenges to the previous enclave for execution, checkpoint the attestation proofs with regard to prediction correctness, and verify the proofs.
Once the auditing enclave catches an incorrect execution, it can generate an alert report, along with an attestation proof.
Resorting to the auditing enclave, the data owner only needs to verify the auditing enclave's attestation proof to determine whether subsequent predictions are offered by a correct model.
It is also noteworthy that the introduction of the auditing enclave can support third-party auditing.

\subsection{Designing Data Structures for Authentication Layer}\label{sec:structure}
\begin{figure}[t]
\small
\vspace{-5pt}
    \centering
    \includegraphics[width=0.8\linewidth]{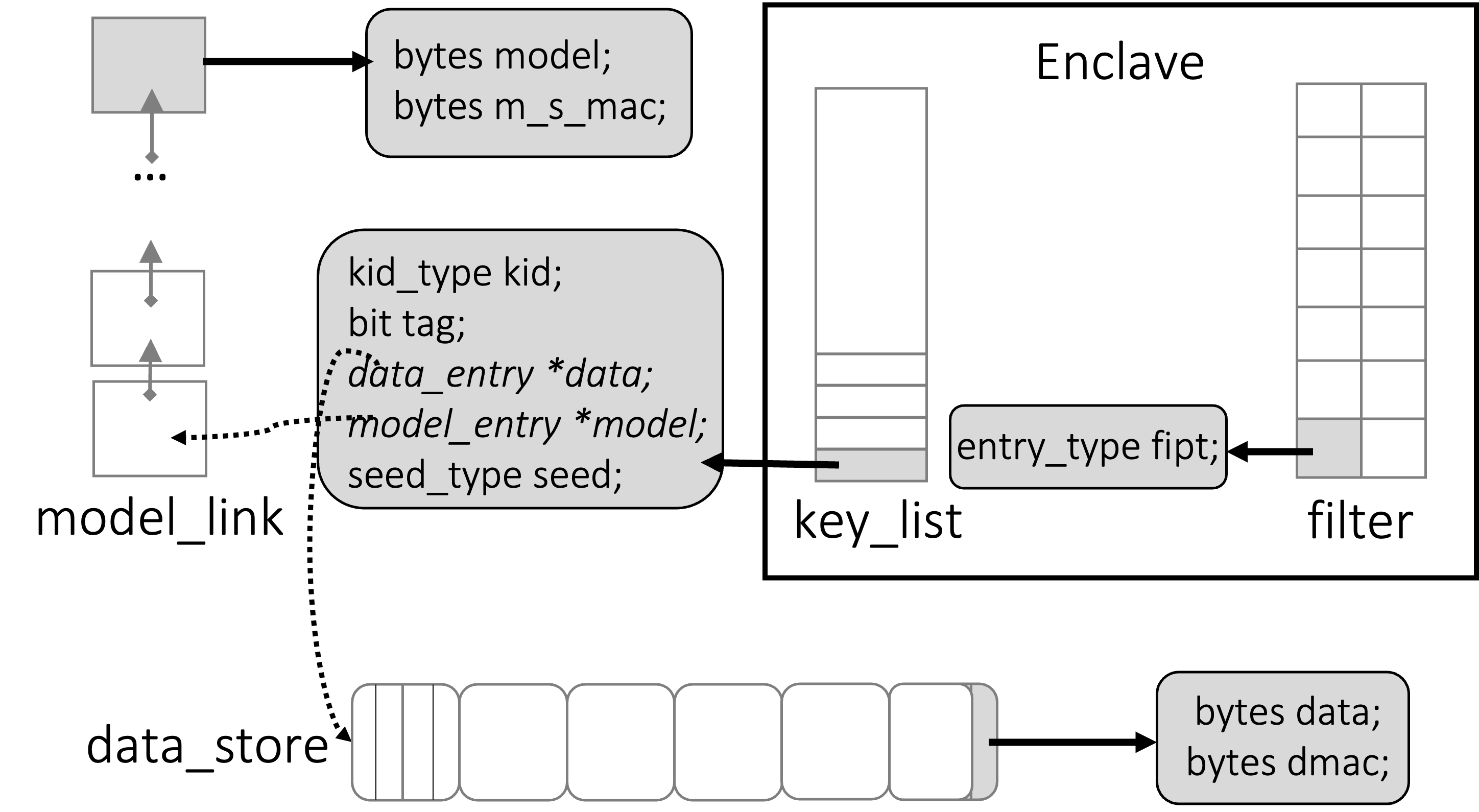}
    \caption{Overview of integrating four data structures.}
    \label{fig:structure}
    \vspace{-20pt}
\end{figure}

To overcome the first challenge, we customize data structures inside and outside the enclave for our logical authentication layer, as Fig.~\ref{fig:structure} shown.
We specifically adopt highly memory-efficient index list and cuckoo filter in the enclave, and use cheap pointers to interlink data slices with correlated submodels out of the enclave.
With the data structures, we are aimed to (\emph{a}) authentically store learning data and submodels out of the enclave, (\emph{b}) efficiently track the lineage from data slices to learned submodels, (\emph{c}) support fast deletion and update on the submodels upon receiving a request of deleting a data point, while maintaining the traceable lineage.

\noindent\textbf{Data Structures} There are two out-enclave structures and two compact in-enclave data structures.
%

\noindent$\bullet$ \textsf{data\_store:} a list for storing sliced learning data and the integrity MACs. 
Its basic unit is respective to a single data point. We note that multiple disjoint data points compose one data slice, as the form of a rounded rectangle in Fig.~\ref{fig:structure}.

\noindent$\bullet$ \textsf{model\_link:} a linked list for storing submodels and the integrity MACs. 
Its basic unit is for one submodel which is learned over the learning data added with one new data slice.
The linkability between two neighboring submodels is enforced by the \textsf{key\_list} introduced later.\looseness=-1

\noindent$\bullet$ \textsf{key\_list:} a list for storing the keys respective to each data point.
Its purpose is to efficiently fetch data and submodels and let data slices and submodels store in a correct order outside the enclave. 

\noindent$\bullet$ \textsf{filter:} a succinct structure for packaging data points while supporting deletion and membership query.
It is operated in the enclave at the very beginning to authenticate data.
It supports fast deletion on a data point, and tells a learning program that this data point is not allowed to be used.\looseness=-1

\noindent\textbf{Detailed Field Description} We now correspondingly describe concrete fields of the \textsf{filter}, \textsf{key\_list}, \textsf{data\_store} and \textsf{model\_link}. We note that each data point is represented as a form of $<$\textsf{kid, data}$>$, where \textsf{kid} is an identifier of the \textsf{data} with a consensual hash function, \emph{e.g.}, non-cryptographic xxHash function.

\textsf{filter}: it is a stateful cuckoo filter residing within the enclave, filled with the fingerprints of the data points. The fingerprints are generated as the form of $\textsf{Enc}(\textsf{kid}||\textsf{data}||\textsf{eid})$, where \textsf{Enc} can be a lightweight encryption algorithm used by \cite{shafi2020secsched}, and \textsf{eid} is the identity of an initialized enclave.
The fingerprints can further be truncated by selecting optimized parameters.
For example, we use $8$~bits per data point for the Purchase dataset.
When a request of deleting a data point is raised, the data fingerprint is removed from the cuckoo filter.
Moreover, the filter helps the enclave in checking if a loaded data point is matched with the respective \textsf{kid}, besides checking its MAC.

\textsf{key\_list}: it maintains a list (or skip list) of keys about the data points, where each key entry has five fields as following. (1) \textsf{kid} serves as the data index.
The orders of the indices enable us to determine the orders of storing and restoring data points (and data slices) plus submodels out of the enclave, and fast fetch them into the enclave in learning and unlearning via pointers. We can see two pointer fields \textsf{*data} and \textsf{*model} later.
(2) \textsf{tag} is a binary field for indicating whether the data point is inserted into or deleted from the cuckoo filter. With the \textsf{tag}, we can avoid repeatedly querying and inserting data points in the filter from scratch, especially in case of deleting a data point from one data slice and relearning over the remaining data.
More importantly, we can use the \textsf{tag} to mitigate \emph{false positives} caused by the existence of certain data point that just shares the same fingerprint with the deleted data point, when looking up the deleted data point later.
(3) \textsf{*data} is used to find the corresponding data point.
(4) \textsf{*model} is used to connect a data slice with the corresponding submodel which is just trained over the training data that just recently adds this data slice. 
Note that one data slice can contain multiple data points, and we let its last data point's \textsf{key} be responsible for linking the corresponding submodel.
(5) \textsf{seed} is prepared for storing the submodels with freshness guarantees.
This seed is generated when a new submodel is built within the enclave, which is filled with a unique and unpredictable randomness.
We assign the fresh seed to the key of the last data point within a data slice for ease of retrieval, resorting to the fact that the key is responsible for linking with the submodel we describe before.

\textsf{data\_store} and \textsf{model\_link}: each entry of the \textsf{data\_store} stores a data point and its MAC \textsf{dmac}, and similarly, each entry of the \textsf{model\_link} stores a submodel and the integrity MAC \textsf{m\_s\_mac}.
We can check a data point's integrity by verifying its MAC and then querying \textsf{filter} with $\textsf{Enc}(\textsf{kid}||\textsf{data}||\textsf{eid})$.
When resorting a submodel, we check the integrity by retrieving the respective seed and rebuilding its MAC \textsf{m\_s\_mac}, so as to verify the authenticity of the submodel.
We note that the submodels and the integrity MACs can be stored in disks or persistent memory, while during learning certain submodel, its previous submodel can reside in DRAM memories.

\noindent\textbf{Mitigating Integrity Violation Attacks} 
We consider potential threats violating the integrity of out-enclave data and submodels. 
We include the enclave identity within each fingerprint to protect against \emph{forking attacks}, so as to prevent the server from scheduling multiple enclave instances to conduct the same unlearning task. 
We also use the in-enclave filter to protect each data point against \emph{replacing attacks}, \emph{e.g.}, replacing a data point with one not matching a particular \textsf{kid}.
As for model checkpoints, we generate trustful and fresh seeds respective to each newly generated submodel, and associate the in-enclave seeds to their integrity MACs. 
With the enforcement, \emph{relocation attacks} and \emph{rollback attacks}, \emph{e.g.}, using a valid submodel from a different or stale address to replace the real stored one, are mitigated.\looseness=-1

\subsection{Designs on Proving Layer}\label{sec:proving}
We are ready to implement the proving layer, by integrating the previous data structures with SGX enclaves to address the second challenge.
%
We start by using the data structures to make the data and submodels operate in an authenticated manner, and meanwhile, scheduling an execution enclave to enforce the correctness of the learning and prediction computations associated with the data and submodels.
It enables a data owner to assert the correctness of the associated computations on deleting a specific data point, by challenging a newly predictive model and verifying attestation proofs.
Later, we introduce an auditing enclave to monitor the subsequent correctness of prediction services, aiming to relieve the data owner's verification cost.

\begin{figure}[t]
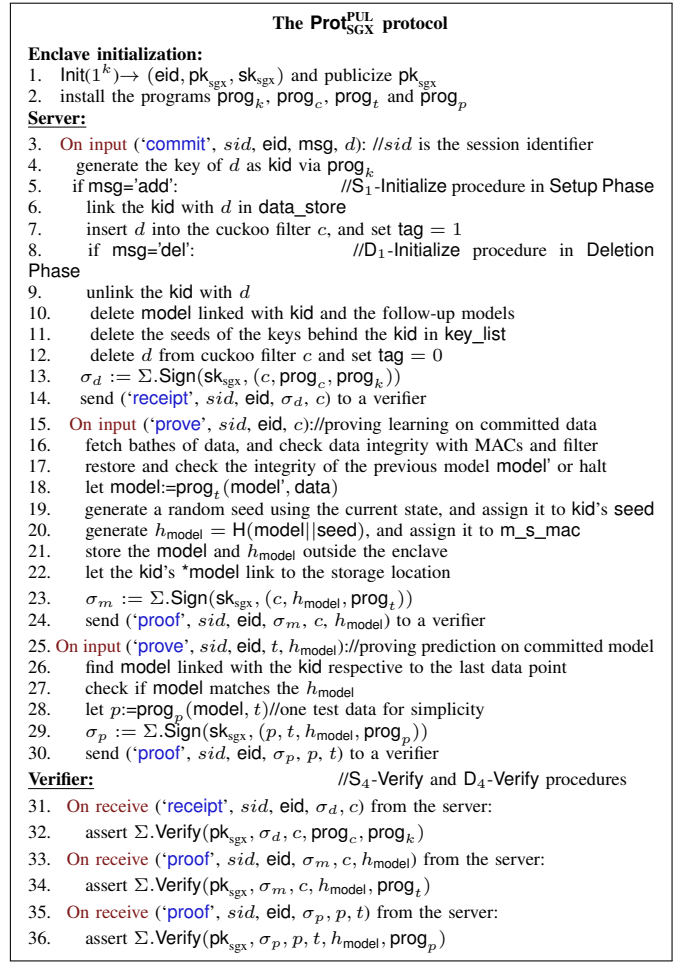

\fbox{
\begin{minipage}[t]{0.94\linewidth}
\scriptsize

\hspace{90pt}
\textbf{The $\textsf{Prot}_{\text{SGX}}^{\text{PUL}}$ protocol}
\vspace{4pt}

\textbf{Enclave initialization:}

\text{1.}
\hspace{2pt}
\textsf{Init}($1^k$)$\rightarrow(\textsf{eid},\textsf{pk}_{\text{sgx}},\textsf{sk}_{\text{sgx}})$ and publicize $\textsf{pk}_{\text{sgx}}$

\text{2.}
\hspace{2pt}
install the programs $\textsf{prog}_{k}$, $\textsf{prog}_{c}$, $\textsf{prog}_{t}$ and $\textsf{prog}_{p}$

\underline{\textbf{Server:}}
\vspace{2pt}

\text{3.}
\hspace{2pt}
\textcolor[rgb]{0.5,0.1,0.1}{On input} (`\textcolor[rgb]{0.1,0.1,0.8}{\textsf{commit}}', $sid$, \textsf{eid}, \textsf{msg}, $d$): //$sid$ is the session identifier

\text{4.}
\hspace{8pt}
generate the key of $d$ as \textsf{kid} via $\textsf{prog}_k$

\text{5.}
\hspace{8pt}
if \textsf{msg='add'}: \hspace{60pt}//$\textsf{S}_1$-\textsf{Initialize} procedure in \textsf{Setup Phase}

\text{6.}
\hspace{12pt}
link the \textsf{kid} with $d$ in \textsf{data\_store}

\text{7.}
\hspace{12pt}
insert $d$ into the cuckoo filter $c$, and set $\textsf{tag}=1$

\text{8.}
\hspace{8pt}
if \textsf{msg='del'}:\hspace{60pt}//$\textsf{D}_1$-\textsf{Initialize} procedure in \textsf{Deletion Phase}

\text{9.}
\hspace{12pt}
unlink the \textsf{kid} with $d$

\text{10.}
\hspace{10pt}
delete \textsf{model} linked with \textsf{kid} and the follow-up models

\text{11.}
\hspace{10pt}
delete the seeds of the keys behind the \textsf{kid} in \textsf{key\_list}

\text{12.}
\hspace{10pt}
delete $d$ from cuckoo filter $c$ and set $\textsf{tag}=0$

\text{13.}
\hspace{6pt}
$\sigma_d:=\Sigma.\textsf{Sign}(\textsf{sk}_{\text{sgx}}, (c, \textsf{prog}_{c},\textsf{prog}_{k}))$

\text{14.}
\hspace{6pt}
send (`\textcolor[rgb]{0.1,0.1,0.8}{\textsf{receipt}}', $sid$, \textsf{eid}, $\sigma_d$, $c$) to a verifier

\vspace{2pt}
\text{15.}
\hspace{2pt}
\textcolor[rgb]{0.5,0.1,0.1}{On input} (`\textcolor[rgb]{0.1,0.1,0.8}{\textsf{prove}}', $sid$, \textsf{eid}, $c$)://proving learning on committed data

\text{16.}
\hspace{8pt}
fetch bathes of data, and check data integrity with MACs and filter

\text{17.}
\hspace{8pt}
restore and check the integrity of the previous model \textsf{model'} or halt

\text{18.}
\hspace{8pt}
let \textsf{model}:=$\textsf{prog}_t(\textsf{model'},\textsf{data})$

\text{19.}
\hspace{8pt}
generate a random seed using the current state, and assign it to \textsf{kid}'s \textsf{seed}


\text{20.}
\hspace{8pt}
generate $h_{\textsf{model}}=\textsf{H}(\textsf{model}||\textsf{seed})$, and assign it to \textsf{m\_s\_mac}

\text{21.}
\hspace{8pt}
store the \textsf{model} and $h_{\textsf{model}}$ outside the enclave

\text{22.}
\hspace{8pt}
let the \textsf{kid}'s \textsf{*model} link to the storage location

\vspace{2pt}
\text{23.}
\hspace{8pt}
$\sigma_m:=\Sigma.\textsf{Sign}(\textsf{sk}_{\text{sgx}}, (c, h_{\textsf{model}}, \textsf{prog}_{t}))$

\text{24.}
\hspace{8pt}
send (`\textcolor[rgb]{0.1,0.1,0.8}{\textsf{proof}}', $sid$, \textsf{eid}, $\sigma_m$, $c$, $h_{\textsf{model}}$) to a verifier

\vspace{2pt}
\text{25.}
\textcolor[rgb]{0.5,0.1,0.1}{On input} (`\textcolor[rgb]{0.1,0.1,0.8}{\textsf{prove}}', $sid$, \textsf{eid}, $t$, $h_{\textsf{model}}$)://proving prediction on committed model

\text{26.}
\hspace{8pt}
find \textsf{model} linked with the \textsf{kid} respective to the last data point

\text{27.}
\hspace{8pt}
check if \textsf{model} matches the $h_{\textsf{model}}$

\text{28.}
\hspace{8pt}
let $p$:=$\textsf{prog}_p(\textsf{model}, t)$//one test data for simplicity

\text{29.}
\hspace{8pt}
$\sigma_p:=\Sigma.\textsf{Sign}(\textsf{sk}_{\text{sgx}}, (p, t, h_{\textsf{model}}, \textsf{prog}_{p}))$

\text{30.}
\hspace{8pt}
send (`\textcolor[rgb]{0.1,0.1,0.8}{\textsf{proof}}', $sid$, \textsf{eid}, $\sigma_p$, $p$, $t$) to a verifier

\vspace{2pt}
\underline{\textbf{Verifier:}}
\hspace{90pt}//$\textsf{S}_4$-\textsf{Verify} and $\textsf{D}_4$-\textsf{Verify} procedures
\vspace{2pt}

\text{31.}
\hspace{1pt}
\textcolor[rgb]{0.5,0.1,0.1}{On receive} (`\textcolor[rgb]{0.1,0.1,0.8}{\textsf{receipt}}', $sid$, \textsf{eid}, $\sigma_d, c$) from the server:

\vspace{2pt}
\text{32.}
\hspace{8pt}
assert $\Sigma.{\textsf{Verify}}(\textsf{pk}_{\text{sgx}}, \sigma_d, c, \textsf{prog}_{c},\textsf{prog}_{k})$

\vspace{2pt}
\text{33.}
\hspace{1pt}
\textcolor[rgb]{0.5,0.1,0.1}{On receive} (`\textcolor[rgb]{0.1,0.1,0.8}{\textsf{proof}}', $sid$, \textsf{eid}, $\sigma_m, c, h_{\textsf{model}}$) from the server:

\vspace{2pt}
\text{34.}
\hspace{8pt}
assert $\Sigma.{\textsf{Verify}}(\textsf{pk}_{\text{sgx}}, \sigma_m, c, h_{\textsf{model}},\textsf{prog}_{t})$

\vspace{2pt}
\text{35.}
\hspace{1pt}
\textcolor[rgb]{0.5,0.1,0.1}{On receive} (`\textcolor[rgb]{0.1,0.1,0.8}{\textsf{proof}}', $sid$, \textsf{eid}, $\sigma_p, p, t$) from the server: 

\vspace{2pt}
\text{36.}
\hspace{8pt}
assert $\Sigma.{\textsf{Verify}}(\textsf{pk}_{\text{sgx}}, \sigma_p, p, t, h_{\textsf{model}},\textsf{prog}_{p})$

\end{minipage}
}

\caption{The $\textsf{Prot}_{\text{SGX}}^{\text{PUL}}$ protocol.}\label{fig:AP_SGX_Prot}
\vspace{-16pt}
\end{figure}
\noindent\textbf{The $\textsf{Prot}_{\text{SGX}}^{\text{PUL}}$ Protocol} From a commit-and-prove perspective, we leverage an enclave to ensure that learning processes are fed with the learning data consistent with pre-committed data as a form of in-enclave \textsf{filter}.
Similarly, we enforce that the later prediction process uses a newly yielded submodel matching the lastly committed submodel in \textsf{model}\_\textsf{link}.
Next, in Fig.~\ref{fig:AP_SGX_Prot}, we elaborate the protocol based on Tram{\`e}r \emph{et al.}'s ``commit-and-prove" functionality~\cite{tramer2017sealed}, which involves both the learning and prediction processes, deducing an implementation for our PoUL defined in Section~\ref{sec:definition}.

To start with, the enclave initializes a pair of public verification key and signing key $(\textsf{pk}_{\text{sgx}},\textsf{sk}_{\text{sgx}})$, and installs the public and correct programs, including $\textsf{prog}_k$, $\textsf{prog}_c$, $\textsf{prog}_t$ and $\textsf{prog}_p$. 
We note that $\textsf{prog}_k$ specifies how to store the keys of the incrementally feeding data in \textsf{key\_list}.
$\textsf{prog}_c$ defines how to insert the feeding data into a cuckoo filter $c$.
$\textsf{prog}_t$ specifies how to incrementally train intermediate model checkpoints while $\textsf{prog}_p$ defines a model prediction process on given test data.
We also note that the data points in the \textsf{data\_store} are fed into the enclave in an incremental manner.

%

\textbf{\emph{(1)~Setup Phase.}} With a collection of data points $D$, the server packs the data into a filter $c$ and build a key list, as shown in line 3-7. 
After that, the data points are unloaded to release the in-enclave memory.
For executing a learning process (see line 15-24), $\textsf{prog}_t$ can appoint batches of data via their keys, and fetch them into the enclave with integrity checking.
The yielded submodels are stored orderly with integrity and freshness guarantees in \textsf{model}\_\textsf{link}, and the last one will be a predictive model, say $h_{\text{model}}$.
Later, the data owner can enjoy the model prediction by offering test data, refer to line 25-30.\looseness=-1

\textbf{\emph{(2)~Deletion Phase}.} Upon deleting one data point $d_u\in D$, the enclave executes the deletion operations on the filter to delete the data point, and meanwhile, invalidates the affected submodels, as demonstrated in line 8-14.
Later, a sequential of new submodels are relearned on the remaining data by executing the steps in line 15-24, and the last submodel is queried by new test data.
The execution processes are similar to that in the setup phase, with the main difference that the new model $h_{\textsf{model}}^{'}$ does not use the data point $d_u$. 

\noindent\textbf{Verifying Correctness Properties} With the $\textsf{Prot}_{\text{SGX}}^{\text{PUL}}$ protocol, the data owner can assert whether the server offers a fake target model, or executes an incorrect unlearning process, or uses an incorrect unlearned model for new predictions, with respect to the deletion on $d_u\in D$. 
The correctness properties are verified based on some necessary assumptions: 
the SGX enclave guarantees integrity and partial confidentiality (\emph{i.e.}, sealing randomness and secret keys); the associated signature scheme $\Sigma$  satisfies unforgeable under chosen message attacks (EU-CMA) security; the hash functions used for generating the enclave identity and integrity MACs are collision-resistant; lastly, all programs running inside the enclave are bug-free.

\noindent\emph{(1)~Verifying target model correctness.}
The data owner can verify that the predictive model is indeed the incrementally learned model $h_{\text{model}}$ over the data $D$, with the following three steps.
\emph{First}, the data owner can verify the authenticity of the filter $c$ and the key list, namely, correctly packing her data, by asserting the validity of $\sigma_d$ and querying the data point membership $d_u$ with the filter, see line 32.
\emph{Second}, the data owner can assert that the last submodel $h_{\text{model}}$ is incrementally learned over the data existing in the filter $c$, on the promise of the validity of $\sigma_d$, the integrity of on-the-fly batches of data and the previous submodels, and the confidentiality of randomly generated \textsf{seed}.
This step relies on the attestation proof $\sigma_m$ and be verified in line 34.
\emph{Finally}, the data owner asserts that the predictive model is the above submodel $h_{\text{model}}$, by verifying $\sigma_p$ in line 36.\looseness=-1

%

\noindent\emph{(2)~Verifying unlearning correctness.} The data owner firstly can verify the validity of a new attestation proof $\sigma_d^{'}$ on the updated filter and key list after the deletion operations.
Particularly, the deletion operations on the key list are two-fold: 
for the $d_u$'s key entry, the changes include \textsf{tag}$=0$, \textsf{*data} unlinked to $d_u$, \textsf{*model} unlinked to the affected submodel, and \textsf{seed}$=$\textsf{NULL};
for other key entries after the $d_u$'s key entry in the \textsf{key}\_\textsf{list}, the \textsf{*model} fields are unlinked to the submodels, and the \textsf{seed} fields are set null.
Next, all affected submodels are relearned on the remaining data until the last submodel $h_{\text{model}}^{'}$ is obtained, which is enforced by the updated filter and the \textsf{key}\_\textsf{list}.
The data owner secondly verifies that the new last submodel $h_{\text{model}}^{'}$ is incrementally relearned over the data exactly excluding $d_u$, by asserting the corresponding signature-based proof $\sigma_{m}^{'}$ against the updated filter $c^{'}$.
Note that the correctness of all affected submodels prior to $h_{\text{model}}^{'}$ are enforced and verified by the enclave itself, with help of our authentication layer.

\noindent\emph{(3)~Verifying new model correctness.} With the last submodel $h_{\text{model}}^{'}$ which is newly generated, the data owner can determine whether it is used for prediction on her given test data, by verifying the new signature $\sigma_p^{'}$ on given new test data $t^{'}$ and $h_{\text{model}}^{'}$.\looseness=-1

\noindent\textbf{Introducing an Auditing Enclave}
We introduce an additional enclave as a trusted auditor to monitor the correctness of the execution enclave and offer the evidences of incorrect execution for post verification.
It derives from our considerations that making the data owner always assert the correctness of each prediction is impractical.
Besides, the previous designs might not be scalable to support third-party regulators to audit unlearning, if they have no access to the data owners' data.
We note that the auditing enclave can work in a centralized or decentralized manner, inspired by Paccagnella \emph{et al.}'s work~\cite{paccagnella2020custos}.

The auditing enclave can be setup on the server's machine and configured with auditing programs, after the interactions between the data owner and the server.
The auditing programs define the auditing logic, including (1) reading the inputs and outputs of \textsf{ECALL} functions invoked by the execution enclave, (2) interacting with Intel's verification server, (3) logging the verification results, and (4) generating reports in the case of verification failure.
At the beginning, the auditing enclave can run a key exchange protocol with the execution enclave, so as to build a secure communication channel~\cite{anati2013innovative}.
The auditing enclave then is stably scheduled when using models for predictions.
Specifically, upon receiving a data owner's test challenge, the auditing enclave can also load the test data due to the defined logic (1), and then monitor the execution of another execution enclave, and lastly assert whether the final prediction output is yielded by the model as expected, according to the logic (2).
From the view of the data owner (and third-party regulators) can retrieve the verification reports returned by the auditing enclave, resorting to the logging history in logic (3) and (4).\looseness=-1

\section{Implementation and Evaluation}
We implement our SGX-capable PoUL instantiation with the logically authentication layer and proving layer.
%
Based on the implementation, our evaluation answers the questions as follows: 
($\textsf{Q}_1$) how is the additional storage cost for PoUL, standing on the already storage workloads of the SISA framework? 
($\textsf{Q}_2$) how does our authentication layer perform, compared against an MHT-based implementation?
($\textsf{Q}_3$) how are the learning/unlearning time complexity and trained accuracy in the SGX-capable PoUL?
%

\subsection{Implementation and Setup} 
We begin with the implementation idea of the SISA framework, and run the training process over a single shard of data in an SGX enclave. 
We omit other shards' training processes, since they can be securely executed in the same manner, and thus we measure prediction accuracy of a single-shard model.

\noindent\textbf{SGX enclave.} We install SGX SDK of version Linux 2.16 to initialize the SGX enclave environment.
We also utilize the Intel's SGX DNNL Library with version 1.1.1 to bootstrap our training tasks. 
All codes are run in hardware mode.

\noindent\textbf{In-enclave programs.} As we described in Section~\ref{sec:proving}, the programs are implemented in C++ majorly including
(1) $\textsf{prog}_c$ for building a cuckoo filter by using Fan \emph{et al.}'s public library\footnote{https://github.com/efficient/cuckoofilter},
(2) $\textsf{prog}_k$ for generating a list for data keys,
(3) $\textsf{prog}_t$ for implementing learning and unlearning a specific model guided by the SISA framework~\footnote{https://github.com/cleverhans-lab/machine-unlearning},
and (4) $\textsf{prog}_p$ for implementing a specific prediction process by restoring a newly trained model via $\textsf{prog}_t$.

\noindent\textbf{Model and dataset.} 
Following the SISA framework's open source library, we adopt a model with two fully connected (FC) layers and re-implement it in C++.
Each FC layer is followed with an activation layer, and the output layer uses one-hot encoding for two classes.  
The model is trained with a mini-batch stochastic gradient descent (SGD) algorithm, and evaluated over the Purchase~\cite{sakar2019real} dataset.
The Purchase dataset is divided into a training set with $280367$ data points and a test set with $31152$ data points.

\noindent\textbf{Setup configuration.} 
We configure the setup information about training the model on the Purchase dataset using the SISA framework.
Concretely, we divide the Purchase dataset into $5$ shards, and each shard contains $56073$ data points.
We then further slice one-shard data points with a range of slice size in $\{1, 3, 6, 12\}$ (see Appendix~\ref{app:exp}).
%
%
For a training case with a fixed slice size, we adaptively select the training parameters, such as the batch size, epoch number and learning rate, leading to a trained model with best accuracy.
%



Besides, our experiments are carried out on an Ubuntu 20.04 server equipped with Intel@ Xeon(R) CPU E3-1505M v5 @ 2.80GHz $\times 8$ CPU and 14.6 GiB of RAM.

\subsection{Evaluation}
%

\begin{table}[ht]
\vspace{-10pt}
\small
 \centering
 \caption{Storage size of customized data structures.}\label{tab:storage} 
 \begin{tabular}{lcccc} 
  \toprule 
  \textbf{Dataset} & \textbf{\textsf{filter}} & \textbf{\textsf{key\_list}} & \textbf{\textsf{data\_store}} & \textbf{\textsf{model\_link}} \\
  \midrule 
  Purchase & $192$~KB & $2.85$~MB & $133.17$~MB & $1.81$~MB\\
  \bottomrule 
 \end{tabular} 
 \vspace{-10pt}
\end{table}
Firstly, we measure the storage costs of our customized data structures, when building the SISA training framework over the Purchase dataset.
We select appropriate parameters, such as the size of each item in the filter and the size of each hidden layer in a model, which are the main impact factors for the storage costs. 
On the Purchase dataset with $56073$ data points in one shard, we let the filter have $2^{16}$ buckets and the fingerprint size of each item be $12$ bits, such that the false positive rate (FPR) for membership query is low enough, nearly $0.003$.
As a result, the filter filled with the one-shard data leads to $192$~KB, as presented in Table~\ref{tab:storage}.
We can also lower the fingerprint size till $8$ bits to obtain a smaller-size filter, but we suffer from a higher FPR value, about $0.041$.
Next, the \textsf{key\_list} is in about $2.85$~MB size, in which each entry needs $416$ bits.
The last two storage sizes are majorly dominated by the data scale (\emph{i.e.}, $600$ nodes in the input layer) and the number of nodes in the FC layer (\emph{i.e.}, $128$ nodes), on the case of using precision training with $32$-bit floating point numbers.
They are the storage workloads already needed by the SISA training.
%
%
Therefore, our authentication layer additionally incurs around $3.00$~MB storage workloads for training on the Purchase dataset, answering the $\textsf{Q}_1$ question.

\begin{table}[htbp]
\vspace{-10pt}
  \centering
  \caption{Performance comparison with an MHT (\textmu s).}
    \begin{tabular}{cccc}
    \toprule
     \textbf{Structures} & \textbf{Insertion} & \textbf{Query} & \textbf{Deletion} \\
    \midrule
     Cuckoo filter & 0.034 & 0.035 & 0.037\\
     MHT &  6.300  & 12.600 & 12.600 \\
    \bottomrule
    \end{tabular}%
  \label{tab:comparison}%
  \vspace{-10pt}
\end{table}%
Secondly, we make a further effort to evaluate the run-time performance of our authenticated layer, compared to a baseline implementation based on an MHT.
As we introduced before, we setup a cuckoo filter inside the enclave for authenticating one shard of data points and later authentically deleting data points.
Here, we compare the performance of insertion, query and deletion operations with that of an MHT using the SHA-256 hash function, over the Purchase dataset.
For demonstrating a best-case comparison, we let the fingerprint size be only $8$ bits.
To the end, we deduce the time consumption for inserting, querying and deleting one data point  in average over the cuckoo filter and MHT, as summarized in Table~\ref{tab:comparison}.
To respond to the $\textsf{Q}_2$ question, our implementation with respect to one data point can achieve about $185$, $360$ and $340$ times time savings in insertion, query and deletion, respectively.
\begin{figure}[ht]
\vspace{-5pt}
    \centering
    \includegraphics[width=1.0\linewidth]{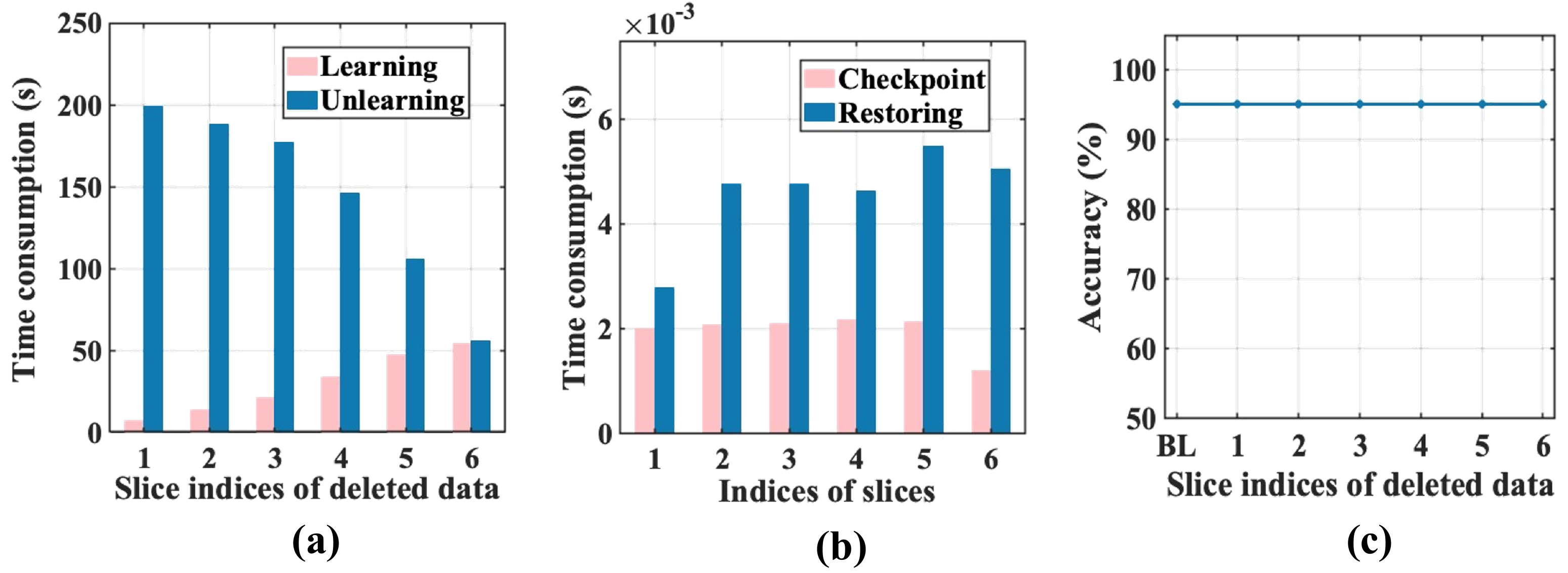}
    \caption{Impact of unlearning one data point.}
    \label{fig:deletion}
    \vspace{-10pt}
\end{figure}

\begin{figure}[ht]
     \centering
     \begin{subfigure}[b]{0.2\textwidth}
         \centering
         \includegraphics[width=\textwidth, height=0.16\textheight]{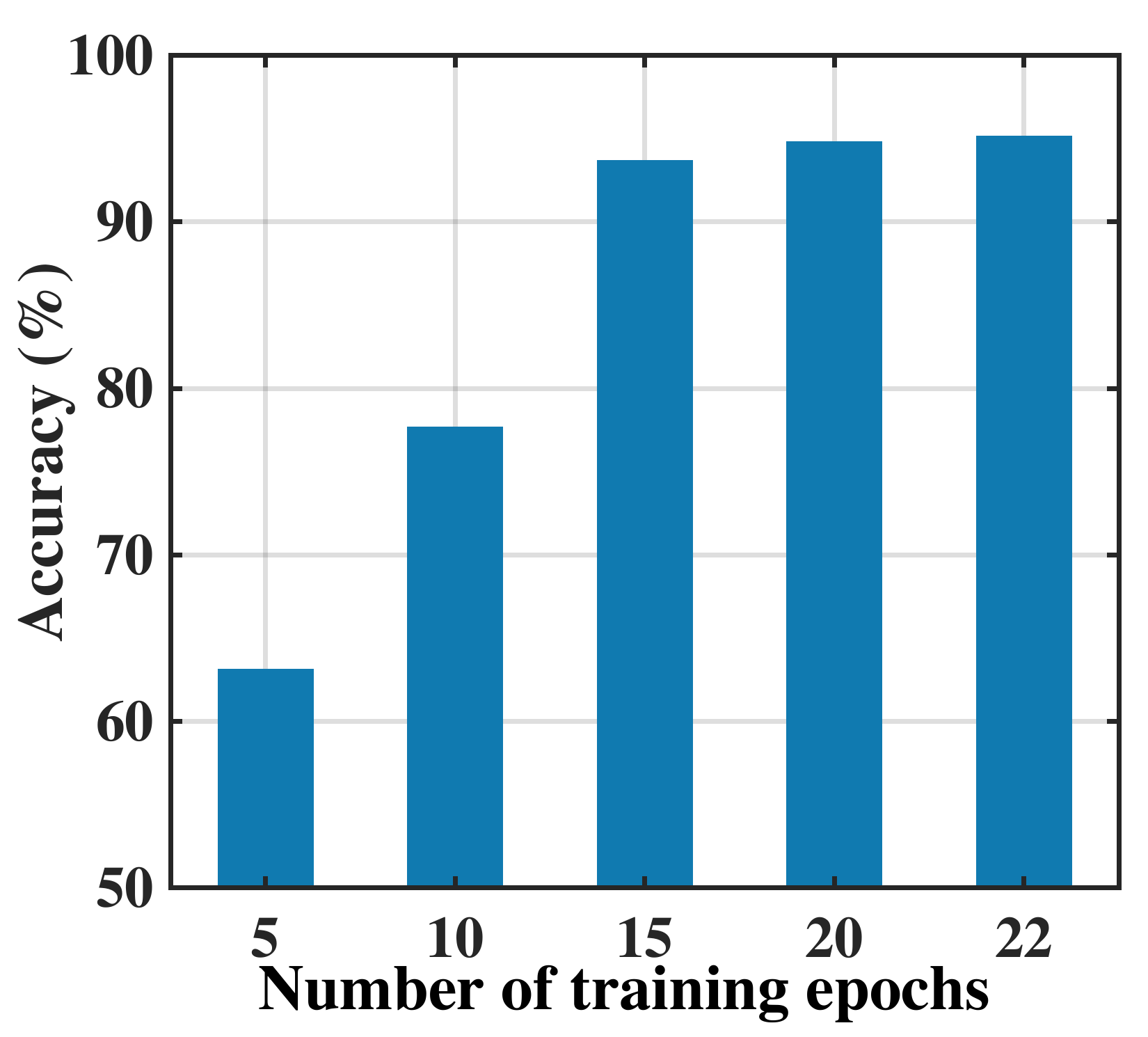}
         \caption{Varying training epochs with fixed batch size}
         \label{fig:epoch}
     \end{subfigure}
     \hspace{2pt}
     \begin{subfigure}[b]{0.2\textwidth}
         \centering
         \includegraphics[width=\textwidth, height=0.16\textheight]{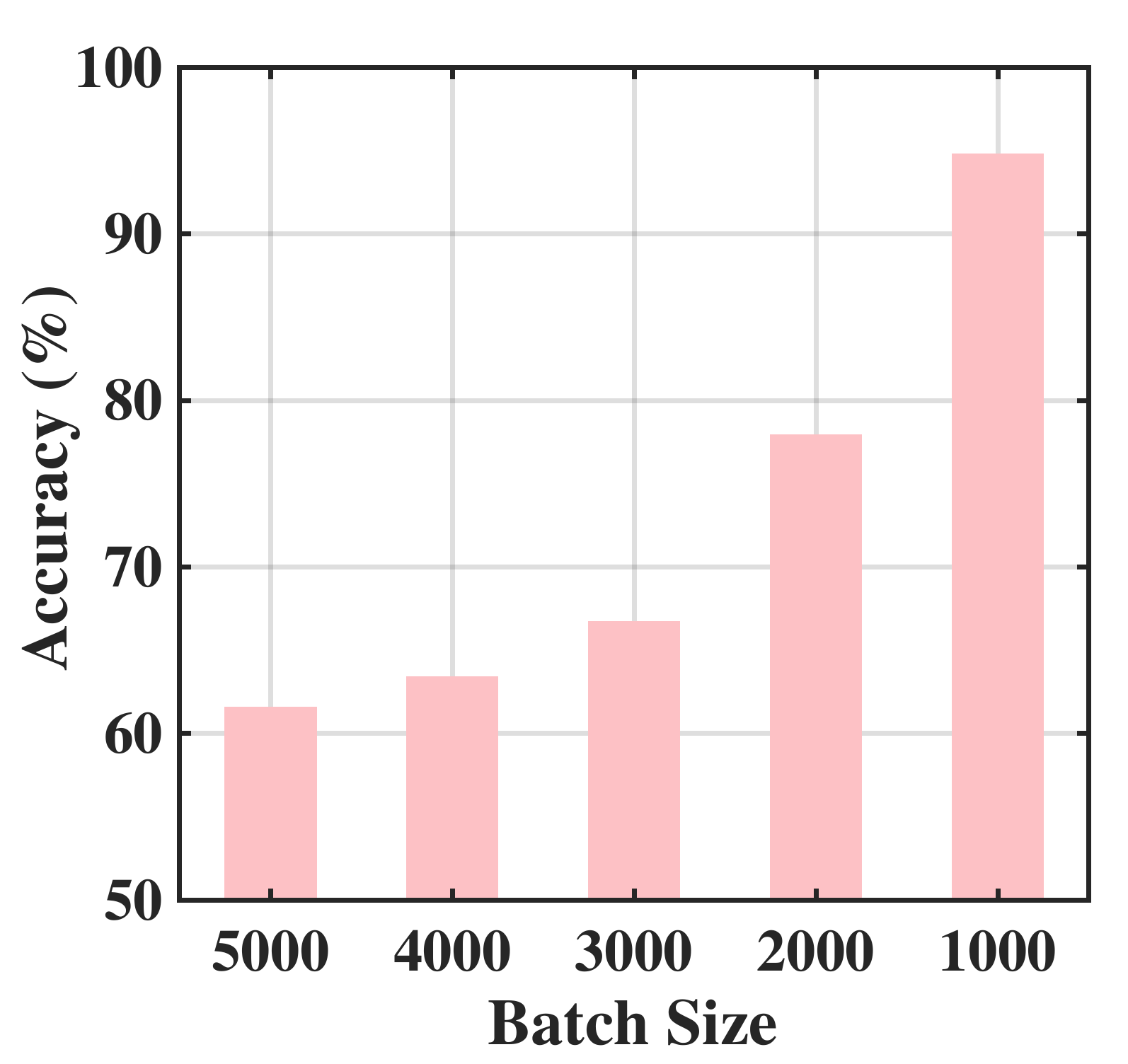}
         \caption{Varying batch sizes with fixed epochs}
         \label{fig:batch}
     \end{subfigure}
     \vspace{-10pt}
        \caption{Accuracy improvement via training parameter tuning.}
        \label{fig:hyperparameters}
\end{figure}

We are ready to evaluate the learning/unlearning time and accuracy in the SGX-capable setting for answering the $\textsf{Q}_3$ question.
On top of our setup configuration with six sequential slices, we measure the unlearning time respective to the positions of the deleted data point, \emph{i.e.}, the slice it falls in.
We also evaluate the individual learning time of each slice before unlearning for comparison.
It is easy to observe from Fig.~\ref{fig:deletion}(a) that learning time turns longer as the number of slices increases, due to that more data need to be trained.
As for unlearning, it consumes more time when the deleted data point falls in the slice with smaller indices, since more impacted submodels need to be retrained.
During the training processes, we notice that model checkpoint and restoring in such an SGX setting take negligible time, compared to the learning/unlearning time, as shown in Fig.~\ref{fig:deletion}(b). 
Besides the time consumption, we also compare the corresponding trained accuracy in Fig.~\ref{fig:deletion}(c).
We use the trained accuracy $95.14\%$ before unlearning as a baseline (shorten as BL in the figure).
We discover from the results that the case of single data point deletion may not influence the original trained accuracy, which might be aligned with a recent study's conclusion~\cite{villaronga2018humans}.
It is noteworthy that due to the memory restriction of the SGX enclave, the trained accuracy mentioned before is obtained by carefully tuning hyperparameters during the SGD training.
Fig.~\ref{fig:hyperparameters} demonstrates the improvement of trained accuracy when (a) the batch size is fixed with $1000$ while the epoch number increases till $22$; and (b) the epoch number is fixed with $20$ while the batch size decreases to $1000$.
As a result, we obtain the baseline accuracy in $95.14\%$ by training with $22$ epochs and a batch size of $1000$.

\section{Analysis of Related Work}
\subsection{Auditing Unlearning}
Auditing unlearning implies that a data owner or third-party regulators can examine if the revoked training data indeed do not exist in an unlearned model residing on a server.
That may relate to a long line of work in auditing the integrity of remotely stored data~\cite{wang2009privacy, wang2010privacy, chen2012robust}, but the focuses here are placed on the model derived from data.
Existing methods~\cite{liu2020have, huang2021mathsf, sommer2020towards, xiao2022verifi} to audit unlearning essentially examine the existence or absence of \emph{certain fingerprint of deleted data} left on an unlearned model, independent of unlearning techniques.
The fingerprint effect either merely relies on original training data~\cite{liu2020have, huang2021mathsf}, or is strengthened via backdoored triggers~\cite{sommer2020towards} and recent watermarking and fingerprinting strategies~\cite{xiao2022verifi}.
The examination is generally performed by means of black-box querying~\cite{liu2020have, huang2021mathsf, sommer2020towards}, aligned with the MLaaS paradigm that offers black-box APIs to users for querying server-side models.
Although the SISA unlearning method can also be audited by letting an auditor re-execute the unlearning process with the submodel checkpoints and the corresponding dataset, the auditing manner 
cannot protect the intellectual property of the submodels, and may not support efficient verification as mentioned in Section~\ref{subsec:goal}.

Different from prior efforts, our work opens a new auditing perspective, in light of a dishonest server who can conduct forging and forking attacks.
For example, the server can forge a model which has an indistinguishable model distribution with that of the target model in question (consider deep neural network models), but the model is never trained on requested data that will be deleted~\cite{thudi2021necessity, shumailov2021manipulating}.
Then, the server forks the target model in question using the forging model, by resorting to the black-box service manner, and claim having deleted data from the forging model while still offering prediction services with the target model.
In such case, neither previous verification metrics via comparing model distribution, nor the above auditing methods via probing fingerprint, can faithfully tell if the requested data is indeed deleted.\looseness=-1

\subsection{TEEs-based Verifiable ML}
%
%
In outsourced ML scenarios, TEEs become a practical tool to enforce computation correctness, either in the training process~\cite{ohrimenko2016oblivious, hunt2018chiron, hynes2018efficient, yuhala2021plinius, zhang2020enabling, ng2021goten, hashemi2021darknight, asvadishirehjini2022ginn} or the inference process~\cite{tramer2018slalom, lee2019occlumency, hanzlik2021mlcapsule, tople2018privado}.
Our work departs from prior implementations merely for inference, but closely relates to the arts, namely~\cite{hunt2018chiron, yuhala2021plinius, zhang2020enabling, hashemi2021darknight, asvadishirehjini2022ginn} for implementing an integrity-protected training (and prediction) pipeline.
Among these arts, very few of them immediately support authentically tracking, storing and updating the lineage of training data in intermediate models, which are particularly required to realize our PoUL.\looseness=-1

\textbf{\emph{Comparison}} TrustFL~\cite{zhang2020enabling} and Plinius~\cite{yuhala2021plinius} are potential to meet our requirements, but they are still not satisfactory.
TrustFL is a TEE+GPUs solution for integrity-preserving federated learning, with the main ideas including (1) individual participants conduct all training iterations on a GPU processor and authentically stores all training intermediate models in alignment with authenticated batches of data; (2) a co-located TEE verifier can randomly re-execute several successive iterations of training for correctness checking after training. 
To authentically store models outside the TEE, it adopts an MHT by packing each intermediate model as a leaf, while preserving the tree root inside the enclave for verification (suppose $n$ leaves in the tree).
When the TEE restores an intermediate model, it needs to compute $O(\text{log}~n)$ hashes along the path from the leaf with respect to this intermediate model to the root, so as to verify the integrity.
Such design is not suitable to realize the PoUL, since we require updating a large number of intermediate models upon receiving a single deletion request, which needs verifying their hashes and computing new hashes in the memory-limited TEE.
As for Plinius~\cite{yuhala2021plinius}, it focuses on storing models in the processor-accessible persistent memory (PM) colocated with a TEE. 
The authors study how to efficiently copy model checkpoints from the secondary storage to the persistent memory, with the eventual aim of making the TEE fast restore the model checkpoints.
Although it does not demand designing efficient data structures to store model checkpoints with dynamic deletion supports, the idea of incorporating efficiently accessible PM is complementary to our work.\looseness=-1
%

\textbf{\emph{Challenging issues of realizing PoUL by combining GPUs with TEEs}}
To make our TEEs-based PoUL instantiation salable to real-life deep neural networks, a suggestive direction is to remove workloads from CPUs-embedded TEEs to GPUs, as promoted by existing works on the training phase~\cite{ng2021goten, hashemi2021darknight, zhang2020enabling, asvadishirehjini2022ginn}; Yet, it is challenging and to be answered by more ongoing rigorous efforts with the previous lessons.
This line of works usually provides probabilistic verification on the integrity of computations delegated to distrustful GPUs by leveraging additional integrity-enhancing algorithms, either delegating all training iterations, \emph{e.g.}, TrustFL~\cite{zhang2020enabling} and GINN~\cite{asvadishirehjini2022ginn} or solely partial linear computations, \emph{e.g.}, Goten~\cite{ng2021goten} and DarKnight~\cite{hashemi2021darknight}.
In addition to integrity-enhancing algorithms, GINN also creatively uses gradient clipping to defend against a fine-gained tampering attack on a single SGD-update step.
Despite the inspiring efforts, following their lessons to realize PoUL needs more exploration, in order to guarantee integrity and practical performance like accuracy, storage overhead and communication complexity caused by frequent system calls.
One big issue is simultaneously ensuring unlearning accuracy and integrity, considering the following two lessons: 
First, Freivalds’ algorithm for integrity checking~\cite{tramer2018slalom, ng2021goten, asvadishirehjini2022ginn} requires carefully mapping the original floating-point numbers into fixed-point ones for maintaining accuracy.
Second, the adoption of gradient clipping demands appropriately adjusting hyperparameters like clipping rate and learning rate for retaining performance, as evaluated by GINN.
Thus, combining GPUs with TEEs for implementing PoUL requires exploring the answer to how the above defences against integrity violation impact unlearning accuracy in the SISA setting.

%
%
%
%
%
%
%

\section{Conclusion and Future Work}
We respond to the need for auditing unlearning in light of the latest attacks on unlearning.
We define the new problem on Proof of Unlearning from the perspective of VC, which is allowed to be realized using various VC techniques.
As an initial effort to push forward auditable unlearning, we propose a native implementation based on Intel SGX.
Standing on top of the initial effort, we collate future work directions as following:
(\emph{i}) \emph{Scaling to specialized hardware accelerators.}
We pursue more efficient auditing in the TEEs setting empowered with specialized hardware accelerators, by incorporating the available optimizing compilers or adapting to the accelerators with the TEEs capability.
(\emph{ii}) \emph{Enabling flexible deletion.}
We seek to explore auditing the compliance of deleting inappropriate sensitive attributes, \emph{e.g.}, race or gender, which can be included in a group of data points, rather than individual data points.
To support deleting the group of data is necessary, as the ML fairness problems become increasingly noticeable (\emph{e.g.}, in face recognition system).
(\emph{iii}) \emph{Protecting data content and privacy.}
To fully respect privacy regulations, auditing unlearning may should stand on the privacy-preserving machine learning/unlearning pipeline, in which data content and data privacy are protected.
Towards this direction, extending our SGX-capable PoUL to work jointly with data-oblivious implementations and differential privacy mechanisms will be explored.

\bibliographystyle{plain}
\bibliography{references}

\appendix

\section{Motivations for our TEEs-Based Solution}\label{app:tec_moti}
We logically combine an authentication layer and a proving layer, and we might implement them by incorporating dynamic ADS with proof-based verifiable computation, to obtain a conceptual solution.
To be specific, ADS enables a user holding the digest of her data, to outsource the data and potential operations on the data to a server, while tracking the integrity of the data.
Proof-based verifiable computation on the ADS can further allow the user to verify the truth of some server-side statement like ``a specific computation is executed on this data matching the ADS, yielding that output", which is faster than reexecution by herself.
Towards the solution, one possibility is combining general-purpose and concise proof systems, such as succinct non-interactive argument of knowledge (SNARK)~\cite{groth2010short}, with dynamic accumulators~\cite{li2007universal, boneh2019batching}, such as RSA-based, pairing-based or Merkle tree-like to initialize the proving layer and the authentication layer, respectively.

But the above incorporation may not be able to handle extensive ML workloads yet, while respecting the underlying computations.
The incorporation might suit the problems with simple computation logic, \emph{e.g.}, linear regression models with dozens of training iterations, but it can become unafforable to handle complex neural network models.
For example, the latest SNARK-based system~\cite{setty2020spartan} takes 86GB memory to prove knowledge of $512\times 512$ matrix multiplication, despite allowing concise proof.
Other memory-efficient proof systems from interactive oracle proofs, ``MPC-in-the-head'' paradigm, garbled circuits and subfield Vector Oblivious Linear Evaluation (sVOLE), have to sacrifice communication efficiency or verification time, as summarized by a recent work~\cite{weng2021wolverine}.
Furthermore, the off-the-shelf proof systems are not designed tailored for training algorithm and usually need modification for adapting to inherent crypto operations~\cite{zhao2021veriml}, \emph{e.g.}, arithmetic operations on finite fields via model quantization.
Recent efforts~\cite{zheng2019helen} also require some changes on the underlying linear model learning algorithms.\looseness=-1

Since the above crypto-assisted solutions are not scalable to generic models, or not compatible to the underlying learning algorithms, or not cost-tolerant yet, our practicality goals cannot be supported.
Therefore, we explore a trusted hardware-based verifiable computation to realize PoUL at as native speed as possible while supporting concise proof and fast verification.\looseness=-1

\section{Functionality Modeling for SGX Enclave}\label{app:fun_sgx}
\begin{figure}[t]
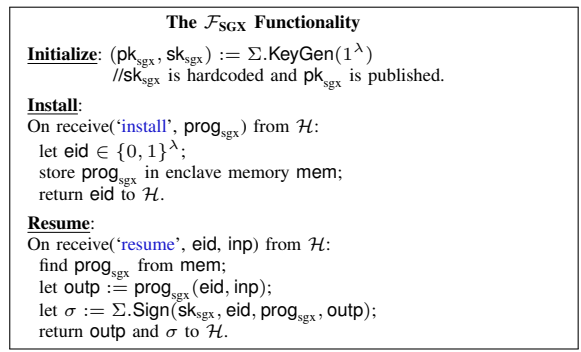

\fbox{
\begin{minipage}[t]{0.8\linewidth}
\scriptsize

\hspace{50pt}
\textbf{The $\mathcal{F}_{\text{SGX}}$ Functionality}
\vspace{4pt}

\noindent\textbf{\underline{Initialize}}:
$(\textsf{pk}_{\text{sgx}}, \textsf{sk}_{\text{sgx}}):=\Sigma.\textsf{KeyGen}(1^{\lambda})$ 

\hspace{30pt}
//$\textsf{sk}_{\text{sgx}}$ is hardcoded and $\textsf{pk}_{\text{sgx}}$ is published.
\vspace{3pt}

\noindent\textbf{\underline{Install}}:

On receive(`\textcolor[rgb]{0.1,0.1,0.8}{install}', $\textsf{prog}_{\text{sgx}}$) from $\mathcal{H}$:

\hspace{2pt}
let \textsf{eid} $\in \{0,1\}^{\lambda}$;

\hspace{2pt}
store $\textsf{prog}_{\text{sgx}}$ in enclave memory \textsf{mem};

\hspace{2pt}
return \textsf{eid} to $\mathcal{H}$.
\vspace{3pt}

\noindent\textbf{\underline{Resume}}:

On receive(`\textcolor[rgb]{0.1,0.1,0.8}{resume}', \textsf{eid}, $\textsf{inp}$) from $\mathcal{H}$:

\hspace{2pt}
find $\textsf{prog}_{\text{sgx}}$ from \textsf{mem};

\hspace{2pt}
let $\textsf{outp}:=\textsf{prog}_{\text{sgx}}(\textsf{eid}, \textsf{inp})$;

\hspace{2pt}
let $\sigma:=\Sigma.\textsf{Sign}(\textsf{sk}_{\text{sgx}}, \textsf{eid}, \textsf{prog}_{\text{sgx}}, \textsf{outp})$;

\hspace{2pt}
return $\textsf{outp}$ and $\sigma$ to $\mathcal{H}$.

\end{minipage}
}
\centering
\caption{The $\mathcal{F}_{\text{SGX}}$ functionality.}\label{fig:sgx}
\vspace{-12pt}
\end{figure}

As depicted in Fig.~\ref{fig:sgx}, there starts with a pair of public verification key and signing key $(\textsf{pk}_{\text{sgx}}, \textsf{sk}_{\text{sgx}})$ at most one time by the Intel's EPID group signature scheme $\Sigma$ on an host $\mathcal{H}$.
Herein, $\textsf{sk}_{\text{sgx}}$ is shielded inside the enclave while $\textsf{pk}_{\text{sgx}}$ is publicly known.
On the activation point \textsf{install}, an enclave can be installed with the program $\textsf{prog}_{\text{sgx}}$, yielding a unique enclave identity \textsf{eid} (a SHA-256 hashsum of the at-launch pages added to the encalve).
Upon receiving a call on the \textsf{resume} activation point with input \textsf{inp}, the program runs on the input and outputs a result.
The result is subsequently attested by the signature scheme, yielding the attestation result $\sigma$.
As a result, the resume result including the output and the attestation result are returned.
The authenticity of attestation results relies on that the EPID signature scheme is existentially unforgeable under chosen message attacks (EU-CMA).

\begin{figure}[t]
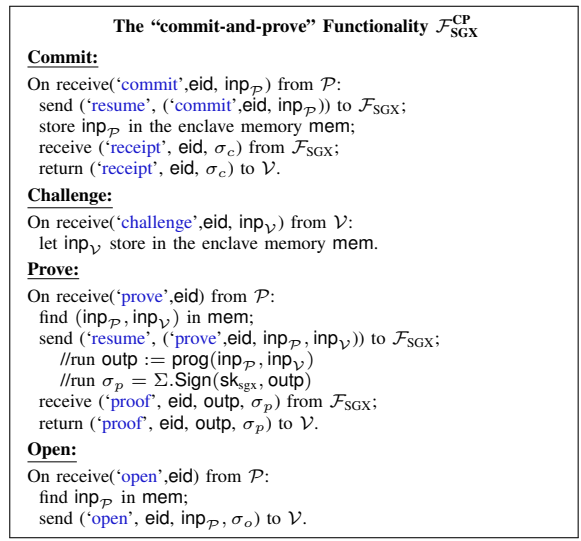

\fbox{
\begin{minipage}[t]{0.8\linewidth}
\scriptsize

\hspace{30pt}
\textbf{The ``commit-and-prove'' Functionality $\mathcal{F}_{\text{SGX}}^{\text{CP}}$}
\vspace{4pt}

\noindent\textbf{\underline{Commit:}}
\vspace{2pt}

On receive(`\textcolor[rgb]{0.1,0.1,0.8}{commit}',\textsf{eid}, $\textsf{inp}_{\mathcal{P}}$) from $\mathcal{P}$:

\hspace{2pt}
send (`\textcolor[rgb]{0.1,0.1,0.8}{resume}', (`\textcolor[rgb]{0.1,0.1,0.8}{commit}',\textsf{eid}, $\textsf{inp}_{\mathcal{P}}$)) to $\mathcal{F}_{\text{SGX}}$;

\hspace{2pt}
store $\textsf{inp}_{\mathcal{P}}$ in the enclave memory \textsf{mem};

\hspace{2pt}
receive (`\textcolor[rgb]{0.1,0.1,0.8}{receipt}', \textsf{eid}, $\sigma_c$) from $\mathcal{F}_{\text{SGX}}$;

\hspace{2pt}
return (`\textcolor[rgb]{0.1,0.1,0.8}{receipt}', \textsf{eid}, $\sigma_c$) to $\mathcal{V}$.

\vspace{2pt}
\noindent\textbf{\underline{Challenge:}}
\vspace{2pt}

On receive(`\textcolor[rgb]{0.1,0.1,0.8}{challenge}',\textsf{eid}, $\textsf{inp}_{\mathcal{V}}$) from $\mathcal{V}$:

\hspace{2pt}
let $\textsf{inp}_{\mathcal{V}}$ store in the enclave memory \textsf{mem}.

\vspace{2pt}
\noindent\textbf{\underline{Prove:}}
\vspace{2pt}

On receive(`\textcolor[rgb]{0.1,0.1,0.8}{prove}',\textsf{eid}) from $\mathcal{P}$:

\hspace{2pt}
find $(\textsf{inp}_{\mathcal{P}}, \textsf{inp}_{\mathcal{V}})$ in \textsf{mem};

\hspace{2pt}
send (`\textcolor[rgb]{0.1,0.1,0.8}{resume}', (`\textcolor[rgb]{0.1,0.1,0.8}{prove}',\textsf{eid}, $\textsf{inp}_{\mathcal{P}}, \textsf{inp}_{\mathcal{V}}$)) to $\mathcal{F}_{\text{SGX}}$;

\hspace{10pt}
//run $\textsf{outp}:=\textsf{prog}(\textsf{inp}_{\mathcal{P}}, \textsf{inp}_{\mathcal{V}})$

\hspace{10pt}
//run $\sigma_p=\Sigma.\textsf{Sign}(\textsf{sk}_{\text{sgx}},\textsf{outp})$

\hspace{2pt}
receive (`\textcolor[rgb]{0.1,0.1,0.8}{proof}', \textsf{eid}, \textsf{outp}, $\sigma_p$) from $\mathcal{F}_{\text{SGX}}$;

\hspace{2pt}
return (`\textcolor[rgb]{0.1,0.1,0.8}{proof}', \textsf{eid}, \textsf{outp}, $\sigma_p$) to $\mathcal{V}$.

\vspace{2pt}
\noindent\textbf{\underline{Open:}}
\vspace{2pt}

On receive(`\textcolor[rgb]{0.1,0.1,0.8}{open}',\textsf{eid}) from $\mathcal{P}$:

\hspace{2pt}
find $\textsf{inp}_{\mathcal{P}}$ in \textsf{mem};

\hspace{2pt}
send (`\textcolor[rgb]{0.1,0.1,0.8}{open}', \textsf{eid}, $\textsf{inp}_{\mathcal{P}}, \sigma_o$) to $\mathcal{V}$.

\end{minipage}
}
\centering
\caption{The ``commit-and-prove'' functionality based on the SGX enclave. We describe the activation points on the side of the prover, and omit that on the verifier side, majorly for verifying attestation reports.}\label{fig:sgx_cp}
\vspace{-10pt}
\end{figure}

With the $\mathcal{F}_{\text{SGX}}$, Tram{\`e}r \emph{et al.}~\cite{tramer2017sealed} define the ``commit-and-prove" functionality in Fig.~\ref{fig:sgx_cp}.
Specifically, the prover is required to firstly commit to an input in the enclave memory.
Upon receiving an unseen challenge from a verifier, the prover executes a previously installed program on the committed input and the verifier's challenge, and yields an attestation proof for the execution correctness.
At a later time, the verifier can also attest the authenticity of the opening of the committed input.

We emphasize that our PoUL problem meets a nature of \emph{knowledge asymmetry} in the above ``commit-and-prove" functionality. For instance, a server can commit to a trained model at first before receiving a data owner's test data as a challenge, and then prove a prediction yielded by a correctly generated model on the test data. 
Notice that we do not require protecting the confidentiality of both the model and data.
Standing on top of the functionality, we need a new design tailored for PoUL satisfying correctness properties and practicality goals as required.\looseness=-1

\section{Discussions}\label{app:discussion}

\noindent\textbf{Supporting Multiple Data Owners}
We consider that the data used to train a predictive model can come from multiple data owners.
In such a scenario, data owners cannot mutually access the data of others, and all of them do not trust the server who trains the model.
To implement PoUL, we start by sequentially authenticating the data owners' data with a unique in-enclave filter in the $\textsf{S}_1$-$\textsf{Initialize}$ procedure.
Suppose the first data owner, with her identity $\textsf{ow}_1$, uploads her data to the server's execution enclave.
The enclave generates the fingerprint of each data point in the form of $\textsf{Enc}(\textsf{kid}||\textsf{data}||\textsf{eid}||\textsf{ow}_1)$. Herein, \textsf{kid} is generated by $\textsf{xxHash}(\textsf{ow}_1||\textsf{data})$.
After traversing her data, the second data owner who prepares for uploading data asserts the current states of the execution enclave, by verifying $(\sigma_d, h_c, \textsf{prog}_c, \textsf{prog}_k)$, where $h_c$ is the digest of the current filter in the enclave.
The above processes repeat until the last data owner's data is traversed, and the filter at the current time packs all data owners' data.
Next, suppose a \textsf{Deletion} phase is invoked for complying with one data owner's deletion request which contains the corresponding \textsf{kid}s of the data point to be deleted.
The execution enclave can retrieve $\textsf{kid}=\textsf{xxHash}(\textsf{ow}_1||\textsf{data})$, and delete the associated data and submodels.
Notice, we consider that the training data points are likely to be \emph{overlapping} in the multi-owner scenario.
Therefore, both keys and fingerprints should include the unique identity information of data owners for identifiability, as described above.\looseness=-1

\noindent\textbf{Enabling Batch Data Deletion}
To delete data in a batch fashion can be necessary, since proving the desired correctness properties with respect to one data point deletion is already cost-consuming for a service provider.
With the concern, the server might wait for deletion requests from the data owners, and record the corresponding \textsf{kid}, and after a waiting period, he executes a \textsf{Deletion} phase in batch, starting from the \textsf{kid} at the foremost location.
We argue such a processing is reasonable, since there often allows a time period to comply with the deletion requests, \emph{e.g.}, one month regulated by the U.K.'s Information Commissioner’s Office\footnote{https://ico.org.uk/for-organisations/guide-to-data-protection/}.

\section{Complementary Experimental Results}\label{app:exp}
\begin{figure}[ht]
\small
    \centering
    \includegraphics[width=0.3\textwidth]{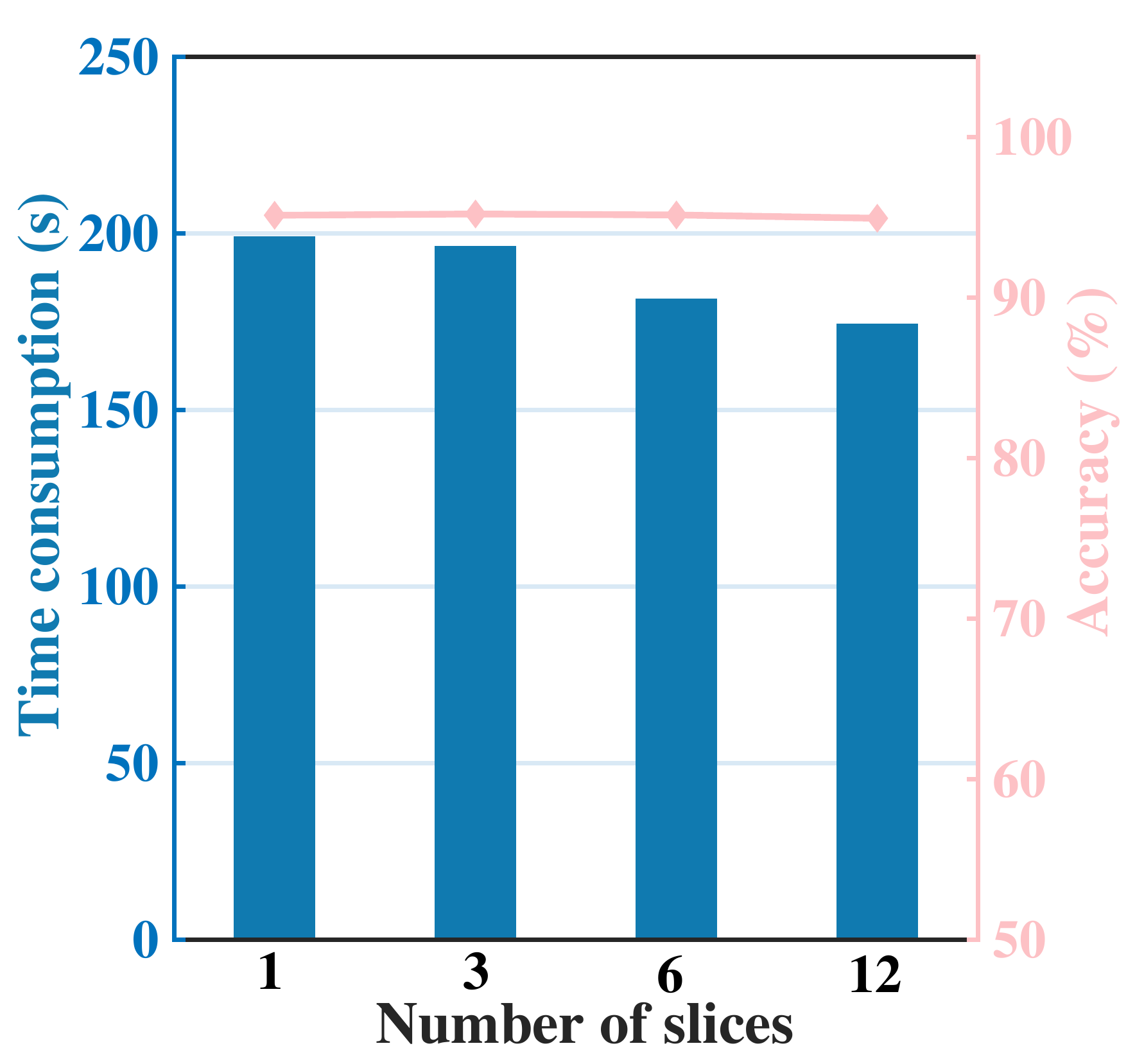}
    \caption{Impact on learning time and accuracy with varying number of slices.}
    \label{fig:slices}
    \vspace{-10pt}
\end{figure}
Recall that we fix the number of slices for our previous evaluation; we now vary the number of slices in one shard, and observe if it impacts the training time and accuracy.
%
%
We divide the one-shard training data into $1, 3, 6$ and $12$ slices, and correspondingly execute the incrementally training processes over them.
As Fig.~\ref{fig:slices} demonstrated, the number of slices makes a really little effect on both the accuracy and the training time.
\end{document}